\documentclass[aip, cha, 
amsmath,amssymb,
floatfix,			
nofootinbib,		
onecolumn,			
reprint,
longbibliography,	
]{revtex4-2}

\usepackage{xcolor}		
\usepackage[total={6.5in,8.75in}, top=1.2in, left=0.9in, includefoot]{geometry}
\usepackage{graphicx}
\usepackage[pdftex,bookmarks, colorlinks, citecolor =blue, breaklinks]{hyperref}
\usepackage{url}
\usepackage{subcaption}

\newcommand{\Eq}[1]{(\ref{eq:#1})}

\newcommand{\Sec}[1]{\S \ref{sec:#1}}
\newcommand{\Fig}[1]{Fig.~\ref{fig:#1}}
\newcommand{\Figs}[2]{Figs.~\ref{fig:#1}-\ref{fig:#2}}
\newcommand{\Tbl}[1]{Table~\ref{tbl:#1}}

\newcommand{\InsertFig}[4]
{\begin{figure}[h!t]
       \centerline{
         \includegraphics[width=#4]{./figures/#1}
       }
       \caption{{\footnotesize  #2}
       \label{fig:#3}}
\end{figure}}

\newcommand{\InsertFigTwo}[5] {
\begin{figure*}[h!t]
       \centerline{
         \includegraphics[width=#5]{./figures/#1}
         \hskip 0.5in
         \includegraphics[width=#5]{./figures/#2}
       }
       \caption{{\footnotesize  #3}
       \label{fig:#4}}
\end{figure*}}

\newcommand{\InsertFigFour}[7] {
\begin{figure*}[h!t]
       \centerline{
\renewcommand{\arraystretch}{0.01}
         \begin{tabular}{cc}
         \includegraphics[width=#7]{./figures/#1}&  \includegraphics[width=#7]{./figures/#2} \\
        \includegraphics[width=#7]{./figures/#3}  &  \includegraphics[width=#7]{./figures/#4}
        \end{tabular}
       }
       \caption{{\footnotesize  #5}
       \label{fig:#6}}
\end{figure*}}

\newcommand{\bN}{{\mathbb{ N}}}

\newcommand{\bR}{{\mathbb{ R}}}

\newcommand{\bT}{{\mathbb{ T}}}

\newcommand{\cB}{{\cal B}}

\newcommand{\cF}{{\cal F}}
\newcommand{\cG}{{\cal G}}

\newcommand{\cN}{{\cal N}}
\newcommand{\cO}{{\cal O}}

\newcommand{\beq}[1]{\begin{equation}\label{eq:#1}}
\newcommand{\eeq}{\end{equation}}

\newenvironment{se}[1]{\equation\label{eq:#1}\aligned}{\endaligned\endequation}
\newcommand{\bsplit}[1]{\begin{se}{#1}}
\newcommand{\esplit}{\end{se}}

\newenvironment{example}[1][]
  {
	\setlength \leftmargini {1.0em}		
	\setlength \topsep {0.5em}			
	\begin{quote}
	{\it Example#1} }
	{\end{quote}
  }
\newcommand{\bexam}[1][:]{\begin{example}[#1]}
\newcommand{\eexam}{\end{example}}

\begin{document}

\title{Recurrence and Stickiness in the Noisy Harper Map}

\author{J.R.~Homan}
\author{J.D.~Meiss} 
\affiliation{Department of Applied Mathematics,
University of Colorado,
Boulder, CO 80309-0526, USA}
\email{Jonathan.Homan@colorado.edu, James.Meiss@colorado.edu}


\date{\today}

\begin{abstract}
\vspace*{1ex}
\noindent

When three types of noise are introduced to the area-preserving Harper map, the Poincar\'e recurrence statistic (PRS) exhibits an extended tail, corresponding to an increased probability of longer recurrence times. For a deterministic case with a mixture of regular and chaotic orbits, regular islands are responsible for a power-law decay in the recurrence distribution. Noise perturbations allow trajectories to access the interior of the islands, and this can enhance their trapping effect, causing many orbits to take longer to return to a neighborhood of their initial conditions and resulting in a slower power-law decay on an intermediate time scale. On a longer time scale, however, the noisy PRS exhibits exponential decay, eventually falling below the deterministic PRS. We compare distributions of trapping and visit times to islands with recurrence times to show the importance of noise in creating tails in the PRS. A simple model of the dynamics---a Markov chain with three states---demonstrates how the slower decay can be caused by noise allowing entry to a previously inaccessible island.

\end{abstract}

\keywords{chaotic transport, area-preserving map, random perturbations, power-law decay}
\maketitle

\begin{quotation}

Chaos is a prominent feature of deterministic dynamical systems and is often thought of as akin to randomness, leading to unpredictability and diffusive motion.  We study the effects of non-deterministic noise on transport in Hamiltonian systems, in particular for an area-preserving map. For the deterministic case, such transport is inhibited by invariant tori. However, in the presence of noise, previously invariant regions of phase space can be accessed. We examine how this affects transport by computing the Poincar\'e recurrence statistic, which measures the fraction of trajectories that have yet to return to a neighborhood of their initial conditions. We study how the trapping within the regular islands surrounding stable periodic orbits depends on the intensity of the noise. Such island chains can inhibit transport when the noise intensity is small, but this effect goes away for sufficiently large noise intensity over long time scales. We show this results in a correlation between trapping time near an island and recurrence time to a well-separated region. A simplified, three-state Markov model elucidates this behavior, showing that a slower decay is introduced when noise is introduced. 

\end{quotation}

\section{Introduction}
The phase space for many dynamical systems is a combination of chaotic and regular regions, typically separated by invariant tori. These can be either absolute barriers when they are co-dimension one or merely impediments to transport when they are of lower dimension (e.g., the phenomena of Arnold diffusion\cite{Arnold64,Lochak92}). When perturbed by noise, such invariant structures are typically destroyed; nevertheless, if the noise is weak enough, these barriers can still be important.

Even when an invariant torus in a deterministic system is destroyed it is often replaced by a partial barrier---a \textit{cantorus}---through which transport can be slow. This phenomena has been mainly studied for area-preserving maps and Poincar\'e sections of two-degree-of-freedom Hamiltonians.\cite{MacKay84, Meiss92a, Meiss15b} Cantori also enclose the island chains surrounding periodic orbits, leading to a power-law decay of survival probabilities,
\[
    P(t) \sim t^{-\gamma},
\]
that is characteristic of Hamiltonian dynamics with mixed regular and chaotic regions. \cite{Karney83, Chirikov84a, Meiss85b, Lange16}  In this paper we study how noise can alter such power laws.

There have been many studies of the effect of noise on chaotic dynamics.
Early papers studied the momentum diffusion coefficient for Chirikov's standard map, 
\cite{Chirikov79b,Rechester80, Karney82}
\bsplit{StdMap}
	x' &= x +y' + \delta x , \\
    y' &= y - \frac{K}{2\pi} \sin(2\pi x) +\delta y , 
\esplit  
on the phase space $M = \bT \times \bR$.
Here, the noise, represented by the perturbation $(\delta x, \delta y)$, is often chosen from a Gaussian distribution with zero mean and standard deviation $\sigma$.
When $k > 2\pi$, \Eq{StdMap} can have stable accelerator modes: elliptic orbits that have linearly growing momenta, $y_t$. In this case, the momentum diffusion coefficient $D$ of the deterministic system diverges. Noise causes $D$ to become finite, since particles can diffuse out of the accelerating islands. Karney et al\cite{Karney82} show that in the presence of noise $D \propto \sigma^{-2}$,
and similar results apply in higher dimensions. \cite{Mezic95}

More recently, the effect of both Gaussian and uniformly distributed noise on transport for \Eq{StdMap} was studied by da Silva et al\cite{daSilva18} setting $\delta x = 0$. They considered $K \approx 4$, where there is essentially only one elliptic region around the fixed point at $(0,0)$, and studied the cumulative recurrence time statistic $P(t)$ (see \Eq{PRS} below) to a chaotic region outside the island. When the noise is non-zero but small ($\sigma \lesssim 10^{-3}$), its presence causes an enhanced trapping over intermediate times before an eventual exponential decay of $P$ at longer times. 
These authors argue that the slower decay of the recurrence statistic is due to trajectories being able to diffuse into islands that were previously inaccessible, causing the Lyapunov exponent of chaotic trajectories to decrease. We will see in \Sec{Recurrence} that the Harper map exhibits similarly enhanced trapping and will relate this enhancement to noise induced trapping of orbits in islands in \Sec{Stickiness}. In the presence of noise, trajectories enter regular islands and can have long orbit segments within the region, increasing the time before recurrence. 

A related phenomenon is the effect of noise on escape for open dynamical systems. Examples include billiards with a ``hole'' in a boundary,\cite{Demers10} naturally unbounded systems like the H\'enon-Heiles model, \cite{Bernal13, Nieto22} and noise-induced escape from potential wells. \cite{Altmann09} For polynomial maps such as the H\'enon quadratic map, noise can cause previously bounded orbits to diffuse to the boundary of an island and then escape to infinity. \cite{Bazzani97} A related case is the effect of noise on chaotic scattering. \cite{Ott85, Altmann10} In particular, noise can allow orbits within an elliptic region to escape (scattering) with an exponentially decaying distribution.\cite{Seone09} For uniformly hyperbolic dynamics, the escape rate is proportional to the (conditionally) invariant measure of the hole. \cite{Ohshika24}

In this paper we study the effect of random perturbations on the Harper map on the torus $M = \bT^2$, given in \Sec{Harper}. Like \Eq{StdMap}, the Harper map exhibits normal and anomalous diffusion due to accelerator modes. \cite{Lebouf98} It has also been studied as an example of a non-twist map, \cite{Shinohara02} and, since it can be thought of as a time periodic composition of two shear flows, as a model for mixing of a passive scalar in a fluid. \cite{Mitchell17}

In \Sec{Harper} we introduce three models for noise: two are Gaussian distributions and one a uniform distribution.
We focus on two parameter choices for the Harper map. One has many regular regions, including rotational invariant circles that separate the phase space. The second, which we refer to as ``nearly ergodic,'' is much more chaotic but still has four visible regular islands.  We will use several notions of recurrence and transit times that are formalized in \Sec{Recurrence}. In \Sec{PRSK06L4}-\Sec{NearlyErgodic} we study the Poincar\'e recurrence distribution to a small region near the hyperbolic fixed point of the map, showing that the addition of noise can, surprisingly, lead to long-time tails on the recurrence distribution. In \Sec{Stickiness} we develop several techniques for detecting the importance of islands on these distributions. 
Finally in \Sec{ThreeState}, we propose a simple three-state Markov model, where one state represents an invariant region of the deterministic case, to elucidate the noise induced tails.

\section{Harper Map Model}\label{sec:Harper}
In this paper we will consider noisy perturbations to the well-known Harper map, \cite{Karney79, Artuso94, Liu94, Saito97, Lebouf98, Shinohara02}
$H: \bT^2 \to \bT^2$, defined by
\bsplit{HarperMap}
    y' &= y + \frac{K}{2\pi} \sin(2\pi x) ,\\
	x' &= x - \frac{L}{2\pi} \sin(2\pi y') .
\esplit
Two typical phase portraits for this deterministic case are shown in \Fig{phase_portraits}.
For panel (a), where $(K,L)=(0.6,4)$, the phase space is divided into two major components by two bands of rotational invariant circles (each surrounds a twistless invariant circle\cite{Shinohara02}). There are also many island chains; the largest surround the fixed points at $(0,0)$ and $(\tfrac12, \tfrac12)$. In \Fig{phase_portraits}(b), with $(K,L)=(1,5)$, the dynamics appears to be more ``nearly ergodic." There are now only four visible islands that surround a pair of elliptic period-two orbits: the highest and lowest islands (red in \Fig{phase_portraits}) comprise one chain, and the pair near $(\tfrac12, \tfrac12)$ make up another. These arise from period doubling bifurcations of the fixed points. However, note that even though all the rotational invariant circles in this phase portrait have been destroyed, they have been replaced by cantori, some of which still have relatively low flux. The result is that the central region, $0.2 \lesssim y \lesssim 0.7$, has fewer points along the orbits shown than its complement. This occurs because we chose all of the initial conditions in the box labeled $\Lambda$ in the figure; these  trajectories do eventually cross the cantori separating the two regions, but this occurs only after many iterates.

\InsertFigTwo{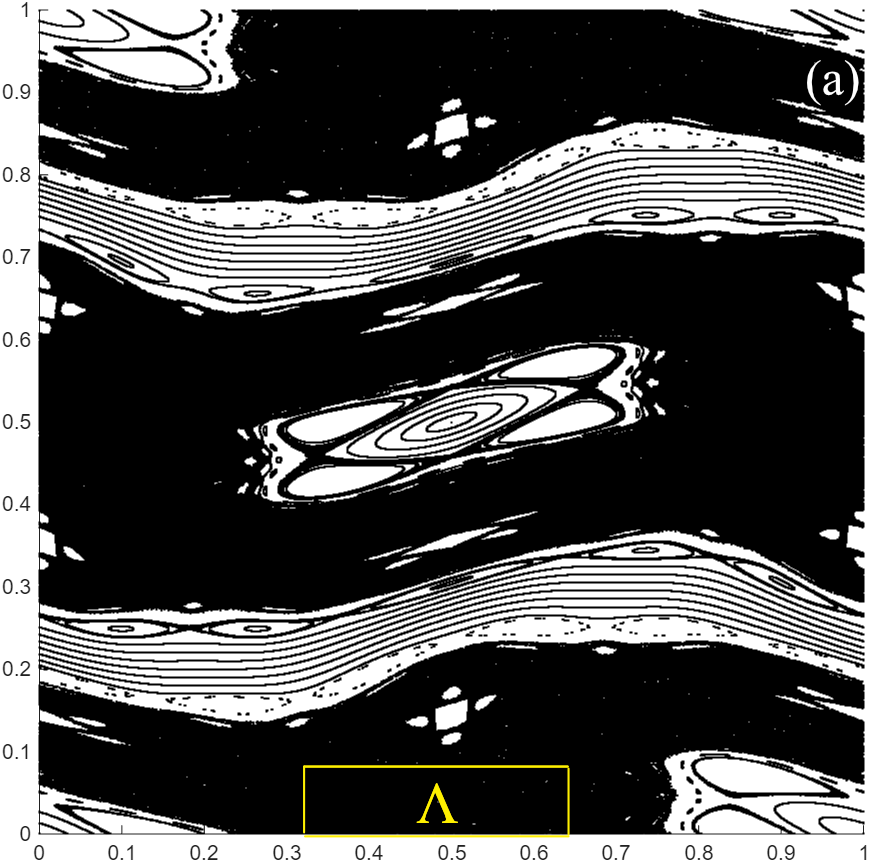}{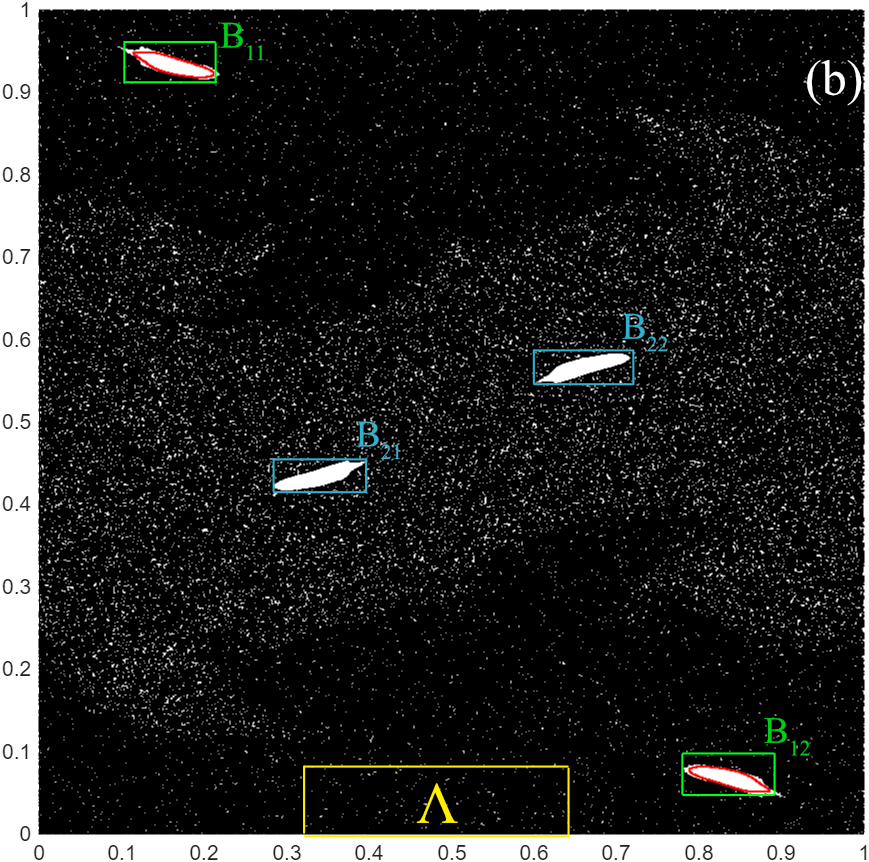}
{Phase portraits for the Harper map \Eq{HarperMap} with (a) $(K,L) = (0.6,4)$, and (b) $(K,L) = (1,5)$. The boxes $\Lambda$, \Eq{StartBox} (yellow),  $\cB_{11}$, $\cB_{12}$, \Eq{IslandBoxes} (green), and $\cB_{21}$, $\cB_{22}$ \Eq{SecondIsland} (blue) used in \Sec{Recurrence} and \Sec{Stickiness} are also shown. For (a), there are $100$ initial conditions on the line $x_0 = y_0$ and each is iterated $t = 3(10)^4$ steps. For (b), the $200$ initial conditions are in the box $\Lambda$ close to the fixed point $(0.5,0)$, on the line segment $0.4 \leq x_0 \leq 0.6$, $y_0=0$ iterated to $t = 3000$. A period-two island chain is identified in red.}
{phase_portraits}{3in}

Adding a noise perturbation to \Eq{HarperMap} results in the map $f_\sigma: \bT^2 \to \bT^2$,
\bsplit{HarperMap_noise}
    y' &= y + \frac{K}{2\pi} \sin(2\pi x) + \delta y ,\\
	x' &= x - \frac{L}{2\pi} \sin(2\pi y') + \delta x ,
\esplit
where $(\delta x,\delta y)$ are random perturbations. Note that the noise in \Eq{HarperMap_noise} is added first to the $y$-component, and this value is then used in computing $x$. Many studies\cite{Karney82, Mezic95} use this formulation. As usual, we denote an orbit of a realization of $f_\sigma$ by a sequence $\{(x_t,y_t), \, t = 0,1,\ldots\}$ such that 
\beq{Orbit}
    (x_{t+1},y_{t+1}) =  f_\sigma(x_t,y_t).
\eeq

We assume that the random forcing in \Eq{HarperMap_noise} has zero mean and standard deviation $\sigma$.  As in da Silva et al\cite{daSilva18}, we consider several cases:

\begin{itemize}
	\item Gaussian noise, i.e., choose independent
    \beq{Gaussian}
         \delta x, \delta y \sim \cN(0, \sigma^2) ,
    \eeq
    with mean zero and variance $\sigma^2$.
	\item Uniform noise on a box with sides $[-R,R]$, i.e., choose independent
    \beq{Uniform}
        \delta x, \delta y \sim \mathcal{U}([-R,R]),
    \eeq
    with variances $\sigma^2 = R/\sqrt{3}$.
    \item Post hoc noise, i.e., added after the deterministic iteration of \Eq{HarperMap},
\beq{PostHoc}
    f_{PH}(x,y)  = H(x,y) + (\delta x, \delta y) .
\eeq
In this case the noise is chosen from a Gaussian distribution \Eq{Gaussian}.
\end{itemize}

An alternative choice of noise that varies the parameters of the system has also been used.\cite{Rodrigues10,Bouchet20} However, we will not consider such phase-space structured noise.

\section{Recurrence Time Statistics}\label{sec:Recurrence}

To characterize transport from a given region $\Lambda \subset \bT^2$
to another $\Omega \subset \bT^2$ we define the \textit{cumulative transition statistic},\cite{Artuso99,Chirikov99,Altmann06b,Cristadoro08}
\beq{TransitTime}
    C(t;\Lambda \to \Omega) =  
    \mathrm{prob}\left(f_\sigma^j(x,y) \notin \Omega, \, \forall j \in [1,t] \,|\, (x,y) \in \Lambda \right) ,
\eeq
i.e., the probability that a trajectory initialized in $\Lambda$ has not yet entered $\Omega$ by time $t$.
Note that $C$ decreases monotonically and that in particular, $C(t) = 1$ for all $t$ if there is an invariant set separating $\Lambda$ and $\Omega$. We numerically estimate \Eq{TransitTime} as
\beq{NumericalC}
    C(t;\Lambda \to \Omega) \approx \frac{N(t)}{N(0)} .
\eeq
Here, $N(0)$  is the number of randomly chosen initial conditions in $\Lambda$, and $N(t)$ 
is the number of those trajectories that have not yet entered $\Omega$ by time $t$. 

For the case $\Omega = \Lambda$, $C$ becomes the \textit{Poincar\'e recurrence time statistic}\footnote
{Also called the recurrence time distribution (RTD) \cite{Altmann06b} or statistic (RTS) \cite{daSilva18}.}
(PRS), the fraction of orbits that have \textbf{not} returned to $\Lambda$ by time $t$:
\beq{PRS}
    P(t; \Lambda) = C(t;\Lambda \to \Lambda) .
\eeq
To compute $P$, we take $N(t)$ as the number of trajectories started in $\Lambda$ that have not reentered $\Lambda$ by time $t$.

It is also convenient to define the \textit{recurrence time}, $t^R : \Lambda \to \bN$:
\beq{RecurrenceTime}
    t^R(x,y; \Lambda) = \min_{ j \ge 1} \{ j \,|\,  f^j_\sigma(x,y) \in \Lambda\} .
\eeq
By the Poincar\'e recurrence theorem, the set of points that never recur, where $t^R = \infty$, has zero measure.

Using the indicator function for the set $B$, $1_B$, we also define the \textit{cumulative trapping time},
$t^T: \bT^2 \times \bN \to \bN$:
\beq{TrappingTime}
    t^T(x,y, k; B) = \sum_{j=0}^k 1_B(f^j_\sigma(x,y)) .
\eeq
Thus $t^T$ is the total time that a trajectory starting at $(x,y)$ spends in $B$ up to time $k$.
For each $(x,y) \in \Lambda$, note that $t^T(x,y,t^R;B) \le t^R(x, y; \Lambda)$.
Finally, we define the \textit{cumulative number of visits} to a region $B$ for a single trajectory,
$n^V: \bT^2 \times \bN \to \bN$:
\beq{VisitTime}
    n^V(x,y,k;B) = \sum_{j=1}^k 1_B(f^j(x,y))1_{\bT^2 \setminus B}(f^{j-1}(x,y)),
\eeq
i.e., the number of times the trajectory segment $\{f^j(x,y)|\, j = 1,\ldots, k\}$ enters $B$ from the outside. Of course $n^V \le t^T$, since a trajectory might remain in $B$ for more iterates after its initial entry to the region.

\subsection{Computing the Poincar\'e Recurrence Statistic}\label{sec:PRSK06L4}
For the examples shown in \Fig{phase_portraits} we will use the box 
\beq{StartBox}
    \Lambda = \{ (x, y)\, | \, 0.32 \leq x \leq 0.64,\, 0 \leq y \leq 0.08\},
\eeq
to compute the cumulative Poincar\'e recurrence statistic \Eq{PRS}. Note that $\Lambda$ is a region near the hyperbolic fixed point at $(0.5,0)$, and---for both cases---appears to contain only chaotic trajectories for the deterministic system. In the following sections, we will use the box \Eq{StartBox} and simply omit $\Lambda$ in our notation: $P(t) = P(t;\Lambda)$. To approximate the PRS, we use \Eq{NumericalC}, with $N(0) = 10^8$ randomly chosen initial conditions uniformly distributed in $\Lambda$, unless otherwise stated.

As a first experiment, we consider the Harper map with parameters $(K,L)=(0.6,4)$ as shown in \Fig{phase_portraits}(a), applying Gaussian noise with standard deviation $\sigma$ \Eq{Gaussian}. The PRS curves for this case are shown in \Fig{prs.K0.6.L4}(a), comparing the deterministic case (dashed black curve) to the noisy cases with eight variances, $10^{-8} \le \sigma^2 \le 10^{-1}$ (solid colors). 

While the PRS curves in \Fig{prs.K0.6.L4}(a) vary considerably in shape as $\sigma^2$ varies, the average recurrence time \Eq{RecurrenceTime} is essentially constant. For the deterministic case, $\langle t^R \rangle = 38.7$ while the average of the noisy cases is almost the same: $\langle t^R \rangle = 39.06$. The fact that the average varies so little with $\sigma$ is consistent with the figure: by $t \approx 200$, more than $90\%$ of the orbits have returned to $\Lambda$ for all cases.

\InsertFigTwo{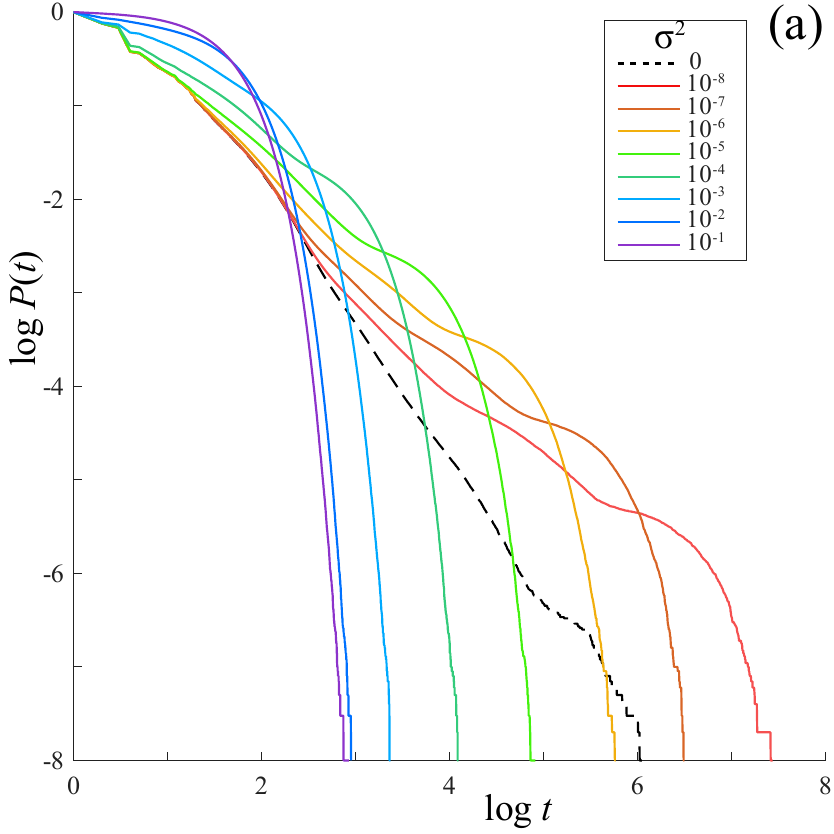}{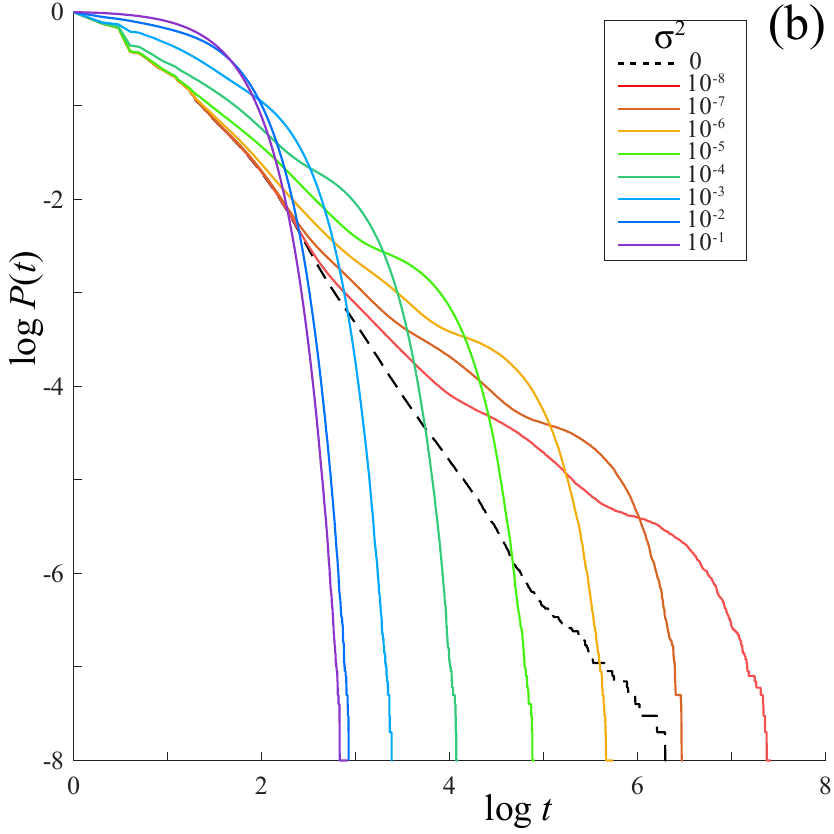}
{Poincar\'e recurrence statistic \Eq{PRS} for the Harper map with $(K,L)=(0.6,4)$: (a) Gaussian noise and (b) uniform noise.
Here we randomly choose $N(0) = 10^8$ initial conditions in $\Lambda$ \Eq{StartBox}. Curves are colored according
to the variance $\sigma^2$. Note that the deterministic cases (dashed) in the two panels differ slightly due to a different random seed for the choice of initial conditions in $\Lambda$.}
{prs.K0.6.L4}{3in}

The deterministic PRS in \Fig{prs.K0.6.L4}  appears to exhibit a power-law decay---a straight line on the log-log plot---with
\beq{PowerLaw}
    P(t;\Lambda) \sim  t^{-\gamma}, \quad \gamma \simeq 1.5587 \pm 0.0003, \quad 10< t < 10^5.
\eeq
Such power laws are characteristic of the stickiness of island chains
and invariant circles in area-preserving maps \cite{Meiss85b,Cristadoro08,Alus17} where a universal exponent $\gamma \simeq 1.56$ is predicted. For larger times, where $P \lesssim 10^{-6}$,
the statistics of our computation become poor, since there are fewer than $100$ nonrecurrent trajectories remaining. Thus, the deviation from the universal power law seen in the figure for $t \gtrsim 10^5$ is a sampling issue.

By contrast, when $\sigma > 0$ the PRS exhibits an eventual exponential decay
\beq{Exponential}
    P(t;\Lambda) \sim e^{-\alpha t} , \; t \gg 1.
\eeq
A similar decay has also been seen for the standard map.\cite{daSilva18}
Notice that when the variance is large, $\sigma^2 \ge 10^{-2}$, exponential decay begins immediately, but when $\sigma^2 < 10^{-2}$, the noisy PRS initially decays as a power law, similar to the deterministic case; this power law is eventually replaced by the exponential \Eq{Exponential}. A similar crossover has been reported for 1D maps.\cite{Floriani95} We report the least squares fits to these rates and the time intervals over which they apply in \Tbl{PRSTable1}.

For the smallest variances, the initial decay of the noisy PRS follows the deterministic case for short times; for example, when $\sigma^2 = 10^{-8}$ they agree up to $t \approx 10^3$. However, as has also been observed by da Silva et al\cite{daSilva18}, the noisy PRS subsequently exhibits slower decay than the deterministic case over intermediate times. For example, when $\sigma^2 = 10^{-6}$ the noisy PRS follows the deterministic one for $0< t < 100$ and then decays as $P(t) \sim t^{-0.879}$ up to $t \approx 4500$, much slower than the deterministic power law \Eq{PowerLaw}. This remarkable, slower decay also occurs for noise levels up to $\sigma^2 \sim 10^{-3}$, although, as shown in \Tbl{PRSTable1}, the power $\gamma$ changes with the variance. For the smallest noise levels, the  PRS has a tail that extends beyond that of the deterministic case, at least up to $t = 10^8$ where our computation stops. For each of the noisy cases the PRS eventually exhibits exponential decay, and when $\sigma^2 \ge 10^{-2}$ there is no observed time interval with power-law behavior.

\begin{table*}[ht]
\begin{centering}
\begin{tabular}{c|cl|cl}
$\sigma^2$	 & $\log t$ range & \multicolumn{1}{c}{$\gamma$}
             &$\log t$ range        & \multicolumn{1}{c}{$\alpha$} \\
\hline
$0$	          &$[1.00,5.00]$ & $1.5587(3)$ &$-$ &$-$  \\
$10^{-8}$     &$[2.85, 5.55]$ &$0.7593(1)$ &$[6.10, 6.79]$ &$2.59374(3) \times 10^{-7}$ \\
$10^{-7}$     &$[2.61, 4.68]$ &$0.7767(2)$ &$[5.09, 6.28]$ &$2.32201(7) \times 10^{-6}$ \\
$10^{-6}$     &$[1.66, 3.65]$ &$0.879(1)$ &$[4.33, 5.56]$ &$2.0110(1)\times 10^{-5}$ \\
$10^{-5}$     &$[0.95, 2.85]$ &$0.942(2)$ &$[3.52, 4.75]$ &$1.6765(1)\times 10^{-4}$ \\
$10^{-4}$     &$[0.85, 2.38]$ &$0.803(5)$ &$[2.82, 3.91]$ &$1.21597(6)\times 10^{-3}$ \\
$10^{-3}$     &$[0.85, 1.74]$ &$0.580(7)$ &$[2.12, 3.26]$ &$6.986(1)\times 10^{-3}$ \\
$10^{-2}$     &$-$ &$-$ &$[1.42, 2.82$] &$2.0054(4)\times 10^{-2}$ \\
$10^{-1}$     &$-$ &$-$ &$[1.42, 2.71]$ &$2.5599(7)\times 10^{-2}$ \\
\end{tabular}
\caption{\footnotesize Slopes of power-law, $\gamma$, and exponential, $\alpha$, fits to the PRS for the Harper map with $(K,L) = (0.6, 4.0)$ and Gaussian noise \Eq{Gaussian}. These were obtained by fitting \Eq{PowerLaw} and \Eq{Exponential} to the curves of \Fig{prs.K0.6.L4}(a) over the time intervals indicated. The error bounds, shown in parentheses for the last digit, represent the least squares errors in the slopes.
The deterministic case has no exponential interval, and the two largest noise levels have no power-law interval.}
\label{tbl:PRSTable1}
\end{centering}
\end{table*}

\subsection{Alternative Noise Distributions}\label{sec:UniformNoise}
To check how the PRS depends on the noise distribution, we now use uniform noise on $[-R,R]$; recall \Eq{Uniform}.
The resulting PRS for $(K,L)=(0.6,4)$ is shown in \Fig{prs.K0.6.L4}(b). These curves are very close to those with normally distributed noise shown panel (a). The differences that do appear are most prominent when $P(t) \lesssim 10^{-6}$, but these are statistical errors for the empirical realization since there are few surviving trajectories.

The three noise varieties of \Sec{Harper} are compared in \Fig{PRS_Compare}. Here, the solid curves represent post hoc noise \Eq{PostHoc}, and the dashed curves are the Gaussian and uniform cases shown previously in \Fig{prs.K0.6.L4}. The biggest differences
occur for the smallest variance $\sigma = 10^{-8}$, and these are
only visible in the figure for $P(t) \lesssim 10^{-5}$ where statistical errors are also expected to affect our results. We can conclude that the three noise methods discussed in \Sec{Harper} are equivalent, at least for the PRS.

\InsertFig{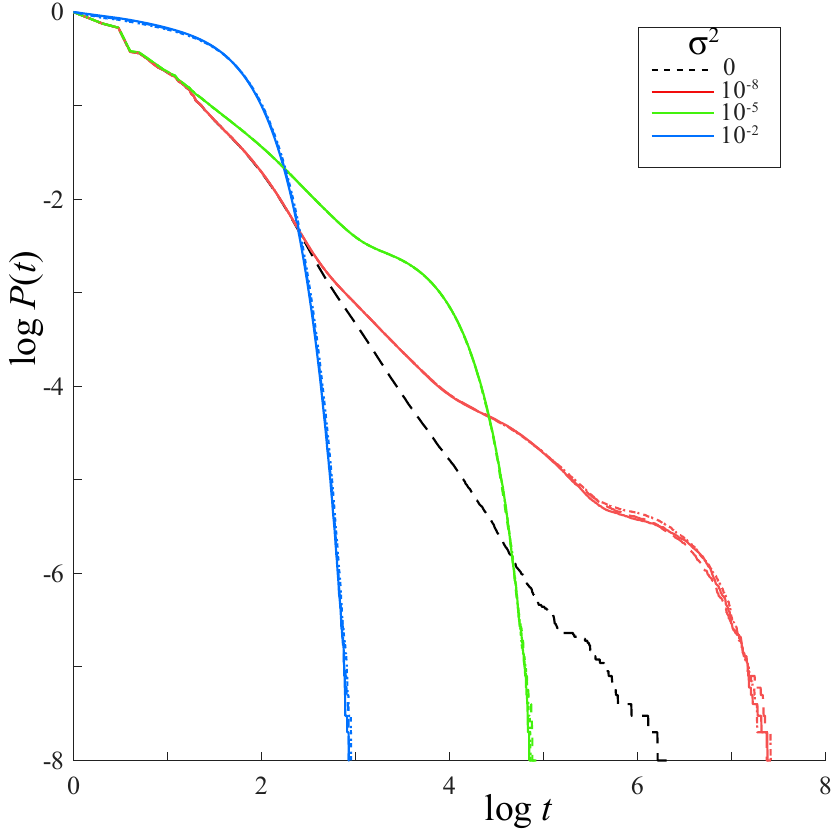}
{Comparison of the PRS for \Eq{HarperMap} with $(K,L) = (0.6,4.0)$ and three types of noise: Gaussian \Eq{Gaussian} (short dashes), uniform \Eq{Uniform} (long dashes) and post hoc \Eq{PostHoc} (solid) for three values of the variance. The curves are almost indistinguishable.}
{PRS_Compare}{3in}

\subsection{Nearly Ergodic Case}\label{sec:NearlyErgodic}
Recurrence time distributions for the nearly ergodic case of \Fig{phase_portraits}(b), where $(K,L)=(1,5)$, are shown in \Fig{prs.K1.L5}. Recall that for these parameters the deterministic map \Eq{HarperMap} has only four visible islands. These are completely inaccessible to orbits in their exterior; nevertheless, they should still exhibit the stickiness that gives rise to the power-law decay \Eq{PowerLaw}.

\InsertFigTwo{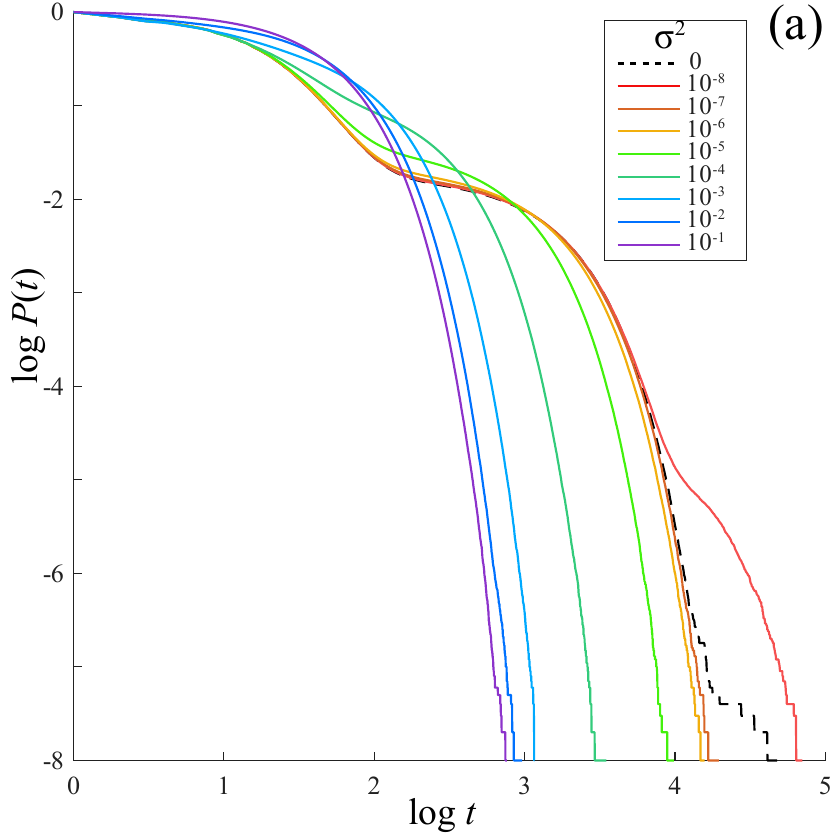}{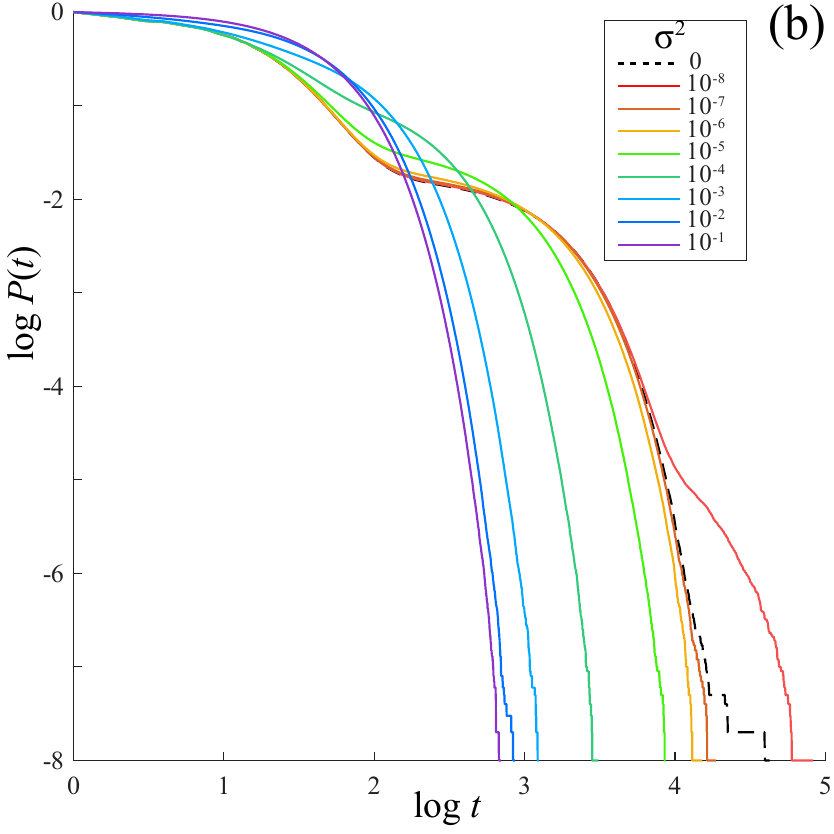}
{Poincar\'e recurrence statistic $P(t;\Lambda)$ for the Harper map \Eq{HarperMap_noise} with $(K,L)=(1,5)$ and $\Lambda$ in \Eq{StartBox}. Both panels use Gaussian noise with variance $\sigma^2$. 
For panel (a) the noise is applied as in \Eq{HarperMap_noise} and, for panel (b) as in \Eq{PostHoc}.}
{prs.K1.L5}{3in}

For \Fig{prs.K1.L5}(a), Gaussian noise is added to the map component by component as in \Eq{HarperMap_noise} and in \Fig{prs.K1.L5}(b) after iteration (post hoc noise) as in \Eq{PostHoc}. The PRS for these two
cases is essentially the same, as we saw for the first set of parameters in \Fig{PRS_Compare}. Again, the difference between the curves is only significant in the tails, where $P(t) \lesssim 10^{-6}$ and statistical errors become significant.

The deterministic dynamics for this nearly ergodic case (black dashed curves in \Fig{prs.K1.L5}) does not show (an obvious) power-law decay over our computation time: it appears to be a combination of two exponentials  \Eq{Exponential}.  Fits to the two coefficients, $\alpha_1$ and $\alpha_2$, and the time intervals over which they apply are given in \Tbl{PRSTable2}. Similar ``hyper-exponential'' distributions have been used by
\citet{Lozej20} to fit recurrence time statistics for the standard map.
Though this PRS should---ultimately---exhibit power-law behavior, seeing this would require many more initial conditions iterated for $t \gg 10^5$.

As we saw in \Sec{PRSK06L4}, the noisy PRS decays more rapidly than the deterministic one when the variance is large, here $\sigma^2 \gtrsim 10^{-5}$. As before, when the variance decreases, there is some indication of a short-time slowdown of the decay, but it is not as prominent as before. For smaller variances the noisy PRS follows the deterministic one for increasingly longer times. For example, when $\sigma^2 = 10^{-7}$, the $P(t)$ curves are almost identical up to $t \approx 2(10)^4$, where $P \approx 10^{-7}$. Note that the rates $\alpha_1$ and $\alpha_2$ for this noise level in \Tbl{PRSTable2} are also close to the deterministic rates.
Though this is still true for $\sigma^2 = 10^{-8}$, in this case $P(t)$ has an extended tail compared to the deterministic case. This corresponds to a third exponential rate (not shown in \Tbl{PRSTable2}) with 
\beq{ThirdAlpha}
    \alpha_3 = 1.1838(3) \times 10^{-4} \mbox{ for } 4.12 < \log t < 4.56 . 
\eeq
Thus on this time scale, as we saw in \Fig{prs.K0.6.L4}, small enough noise reduces the recurrence probability on long time-scales. This---as we will discuss in the next section---is due to the islands.


\begin{table*}[ht]
\begin{centering}
\begin{tabular}{c|cl|cl}
$\sigma^2$	 & $\log t$ range & \multicolumn{1}{c}{$\alpha_1$}
             &$\log t$ range        & \multicolumn{1}{c}{$\alpha_2$} \\
\hline
$0$	          &$[0.00,1.98]$ & $4.67(4) \times 10^{-2}$ &$[2.60, 4.08]$ &$8.5242(6)\times 10^{-4}$  \\
$10^{-8}$     &$[0.70, 1.79]$ &$4.46(5)\times 10^{-2}$ &$[2.47, 3.95]$ &$7.885(5)\times 10^{-4}$ \\
$10^{-7}$     &$[0.30, 1.83]$ &$4.33(5)\times 10^{-2}$ &$[2.77, 4.06]$ &$8.965(1)\times 10^{-4}$ \\
$10^{-6}$     &$[0.30, 1.86]$ &$4.18(5)\times 10^{-2}$ &$[2.62, 4.04]$ &$9.9472(6)\times 10^{-4}$ \\
$10^{-5}$     &$[0.30, 1.75]$ &$4.31(5)\times 10^{-2}$ &$[2.63, 3.79]$ &$1.6872(1)\times 10^{-3}$ \\
$10^{-4}$     &$[0.30, 1.60]$ &$3.93(7)\times 10^{-2}$ &$[2.27, 3.29]$ &$5.4444(7)\times 10^{-3}$ \\
$10^{-3}$     &$[1.78, 2.92]$ &$1.4081(1)\times 10^{-2}$ &$-$ &$-$ \\
$10^{-2}$     &$[1.42, 2.79]$ &$2.1664(3)\times 10^{-2}$ &$-$ &$-$ \\
$10^{-1}$     &$[0.00, 2.72]$ &$2.5584(6)\times 10^{-2}$ &$-$ &$-$ \\
\end{tabular}
\caption{\footnotesize Exponential fits to the PRS for the Harper map with $(K,L) = (1, 5)$ and Gaussian noise \Eq{Gaussian}. These were obtained by fitting \Eq{Exponential} to the curves of \Fig{prs.K1.L5}(a), over the intervals of $\log t$ shown. The error bounds, in parentheses, represent the least squares errors in last digit.  When $\sigma^2 = 10^{-8}$, there is a third exponential on the longest time scales, see \Eq{ThirdAlpha}.}
\label{tbl:PRSTable2}
\end{centering}
\end{table*}

\section{Noise Enhanced Stickiness}\label{sec:Stickiness}
We think of the long tails on $P(t)$ seen in \Figs{prs.K0.6.L4}{prs.K1.L5} for small, nonzero variances as \textit{enhanced stickiness}. One hypothesis that might explain the tails is that in the presence of small noise, trajectories are able to access the interior of islands that are inaccessible in the deterministic case. This is supported in the work of Ohshika et al \cite{Ohshika24} for the standard map, who argued that the effective Lyapunov exponent of a noisy, chaotic trajectory can be reduced from that of the deterministic case when islands can be breached. 
Similar results have been seen for the Josephson map with uniform noise. \cite{Ishizaki00}  This map has accelerator modes, and it was shown that the anomalous diffusion exponent can increase in the presence of noise due to orbits diffusing into these accelerator modes.

To understand this phenomena in more detail, we will study recurrence \Eq{RecurrenceTime}, trapping \Eq{TrappingTime},  and visit \Eq{VisitTime} time for the nearly ergodic case of \Sec{NearlyErgodic}.

\subsection{Trapping near an Island Chain}
Our first goal is to compute the trapping time for the period-two island shown in red in \Fig{phase_portraits}(b). For computational simplicity we enclose the islands in a pair of boxes,
\bsplit{IslandBoxes}
    \cB_{11} &= \{ (x, y) \,|\, 0.1013 \leq x \leq 0.2206,\, 0.9128 \leq y \leq 0.96\}, \\
    \cB_{12} &= \{ (x, y) \,|\, 0.78 \leq x \leq 0.89, 0.047 \leq y \leq 0.085\} ,
\esplit
as sketched in \Fig{phase_portraits}.  We let $\cB_1 = \cB_{11} \cup \cB_{12}$ denote the region associated with this period-two  chain.

As a first visualization of the island trapping,  \Fig{trapping_hist.K1.L5} shows a 2D histogram of trajectories that start in $\Lambda$, comparing the recurrence time $t^R(x,y;\Lambda)$ \Eq{RecurrenceTime} with the cumulative trapping time in the islands $t^T(x,y,k;\cB_1)$ \Eq{TrappingTime}, setting $k = t^R$. To do this, the data for $10^8$ trajectories is partitioned into $200 \times 200$ bins, where the bins are evenly spaced for $(\log (t^R+1), \log (t^T+1)) \in [0,5]\times [0,5]$. The resulting 2D histograms are shown for four noise levels in the figure. The colors represent $n$, the number of trajectories in each bin.

\InsertFigFour{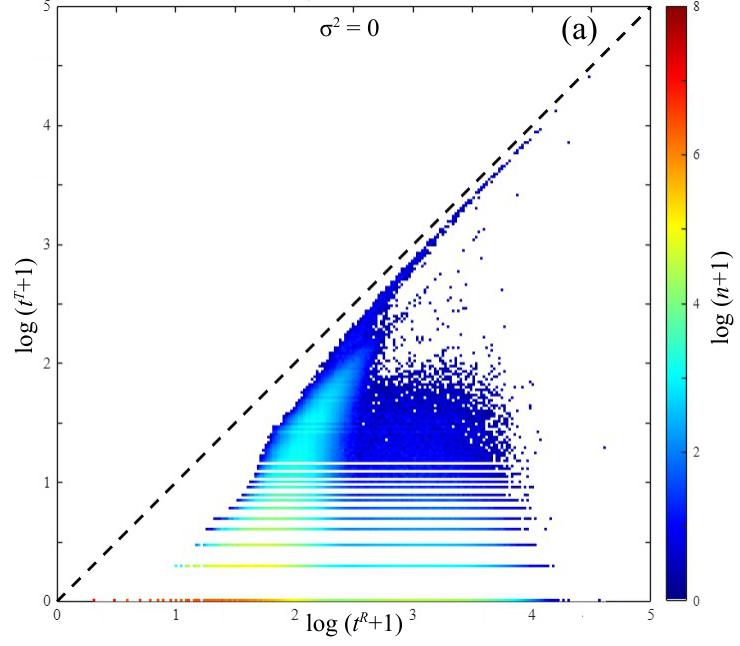}{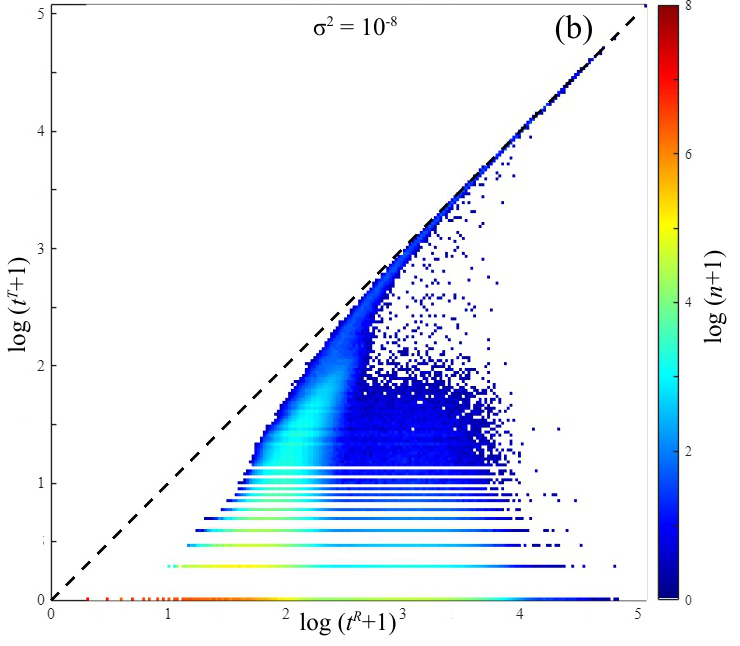}{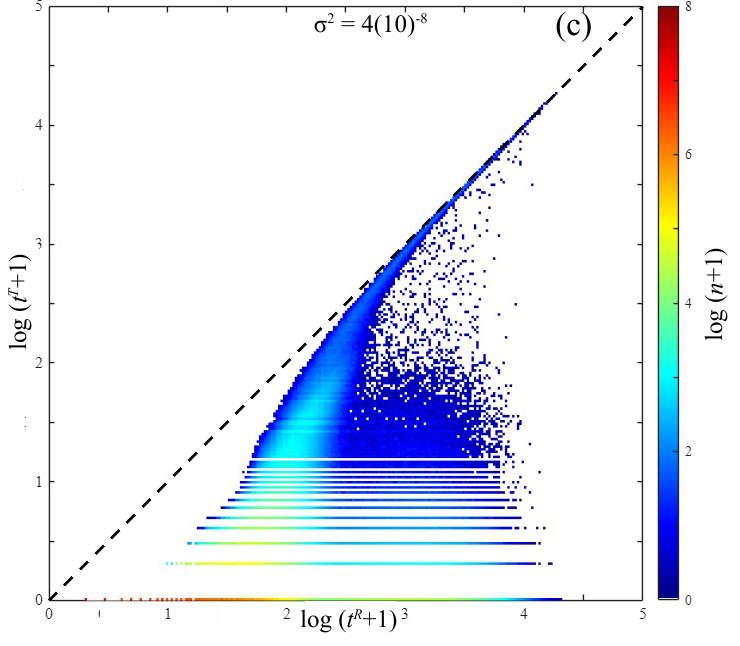}{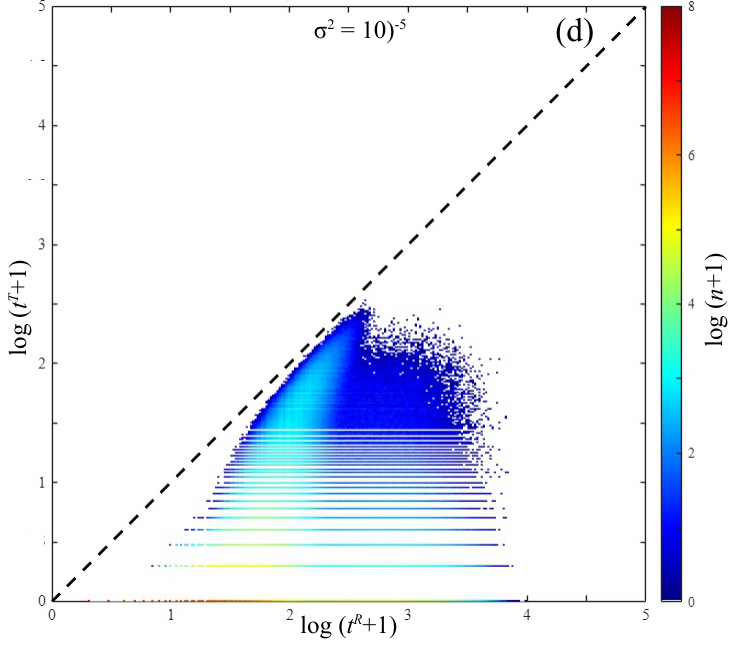}
{Histograms of trapping time, $t^T$ in $\cB_1$ \Eq{IslandBoxes}, versus recurrence time, $t^R$ to $\Lambda$ \Eq{StartBox}, for \Eq{HarperMap_noise} with $(K,L)=(1,5)$. The four panels correspond to variances (a) $\sigma^2 = 0$ (deterministic case), (b) $10^{-8}$, (c) $4(10)^{-8}$, and (d) $10^{-5}$ with Gaussian noise \Eq{Gaussian}. The dashed black line is $t^T = t^R$.}{trapping_hist.K1.L5}{3in}

The axes (including the color bar) are all plotted logarithmically, which is the cause of the striations along the vertical axis for small $t^T$: the trapping time must be an integer. Since each variable in the histogram can be zero, we increment by one so that the logarithm is non-negative. For example, when the population in a bin is zero, we have that $\log(n+1) = 0$, and these bins are colored white. For all noise levels, the maximum number of orbits in any particular bin is $\cO(10^7)$; these are red in the figure. Most of the orbits have recurrence times in the range $1 \le t^R \le 200$, which is consistent with \Fig{prs.K1.L5}, since $P(200) \lesssim 0.015$. The diagonal line (dashed) in each of the plots signifies the maximum trapping time, since $t^T \le t^R$. 

For \Fig{trapping_hist.K1.L5}(b)-(c), where $\sigma^2 \le 4(10)^{-8}$, there is a tail in the histograms for the orbits with the longest recurrence times, 
\[
    t^T \sim t^R, \quad t^R \in [10^3,10^5] .
\]
Trajectories in the tail tend to spend much of their time within the region $\cB_1$. Though this effect is seen for the deterministic case in panel (a), 
\[
    t^T \sim 0.75 t^R , \quad t \in [10^3,10^4] ;
\]
it is weaker and smaller. The stronger tail for small, nonzero $\sigma$ shows the importance of noise in trapping trajectories in the interior of the island chain, which we will discuss further below. 

\subsection{Visits to an Island Chain}
As a second illustration of the importance of the islands in trapping trajectories for the noisy map, we compute the cumulative number of visits $n^V$ \Eq{VisitTime} to the region $\cB_1$ for $N(0) = 10^8$ new realizations of trajectories started in $\Lambda$. A 2D histogram of visit versus trapping time, again with $200 \times 200$ log-bins, is shown in \Fig{visit_hist.K1.L5}. As before, the black dashed line indicates the maximum: $n^V$ cannot be greater than the trapping time. Note that trajectories with $n^V \sim t^T$ should not be considered trapped since they leave $\cB_1$ as often as they visit. 

\InsertFigFour{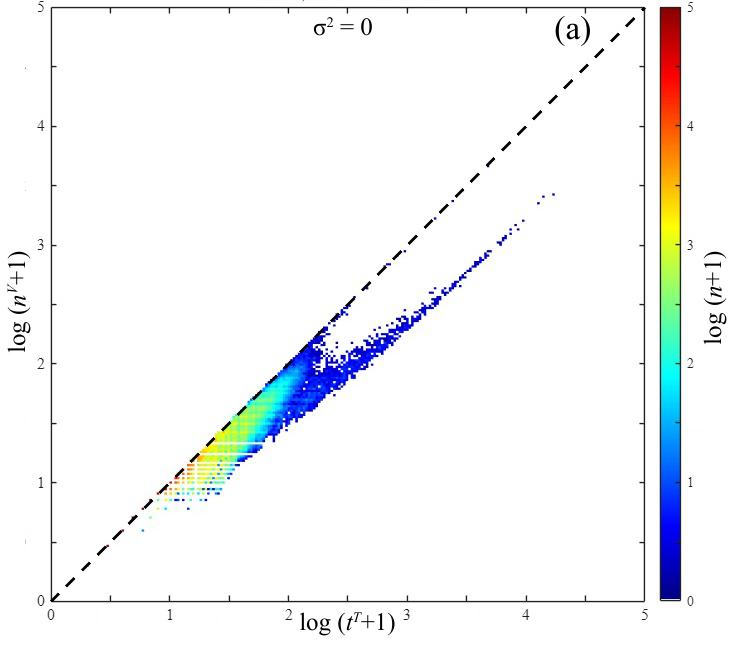}{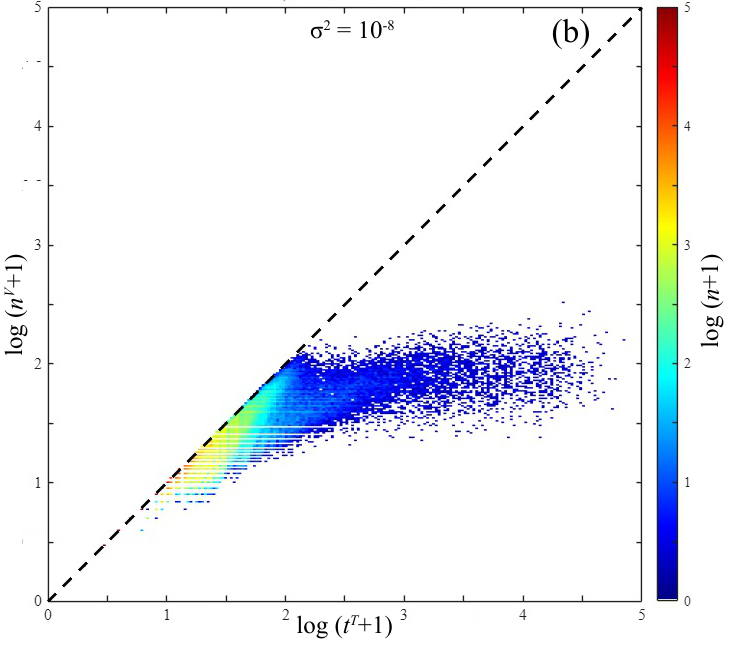}{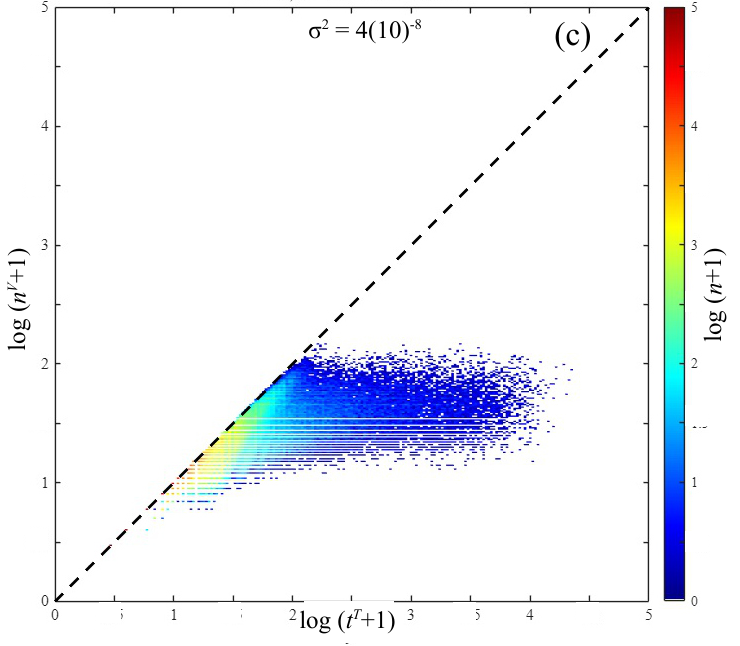}{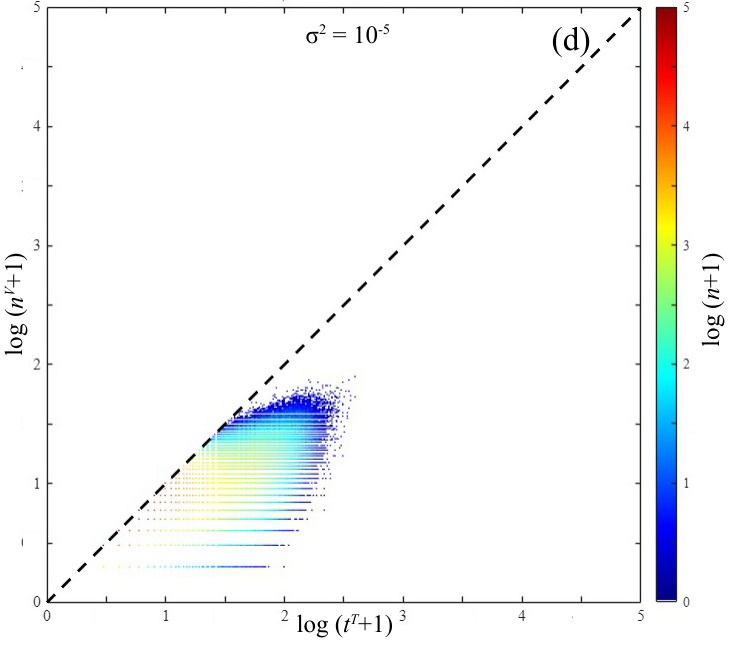}
{Histogram of island chain visit instances, $n^V(x,y,t^R; \cB_1)$ \Eq{VisitTime} as a function of $t^T(x,y,t^R; \cB_1)$ \Eq{RecurrenceTime} for the Harper map \Eq{HarperMap_noise}. The parameters, ranges, and color scale are the same as for \Fig{trapping_hist.K1.L5}.}
{visit_hist.K1.L5}{3in}

For the deterministic case, \Fig{visit_hist.K1.L5}(a), there is a tail in the histogram with
\[
    n^V \sim 0.178 t^T, \quad t^T \in [10^2,10^4].
\]
The implication is that although the longest trapped, deterministic orbits do \textit{enter} $\cB_1$, each such trajectory \textit{remains} in $\cB_1$, on average, for approximately five iterates. In other words, even though an island is ``sticky,'' the probability of becoming entrained by the nested family of cantori with small fluxes near the island is small.
This is also consistent with the fact that we do not see a power law for the decay of $P(t)$ over the time range of \Fig{prs.K1.L5}.

Compared to the deterministic case, the noisy cases have a drastically different distribution for long $t^T$. Trajectories that have a long trapping time now exhibit fewer visits to the island chain, resulting in a nearly flat tail. Though most trajectories do not visit $\cB_1$ as we saw in \Fig{trapping_hist.K1.L5}, the ones that do are trapped for longer periods, as evidenced by their relatively small visit numbers compared to the overall trapping time. Once again, when the intensity of the noise is strong, in panel (d), the maximum trapping time is smaller and the distribution shifts towards the line $n^V = t^T$.

\subsection{Trapping near a Second Island Chain}
In \Fig{trapping_hist.K1.L5}, we saw that there is a population that does not spend much time trapped in the island chain $\cB_1$ yet still exhibits long recurrence times. A possible cause is the influence of the second chain: recall that the nearly ergodic case of \Fig{phase_portraits}(b) has a pair of period-two islands. 
To understand the importance of this second chain we define two new boxes,
\bsplit{SecondIsland}
    \cB_{21} &= \{ (x, y) \,|\, 0.2821 \leq x \leq 0.3937, 0.4149 \leq y \leq 0.4539 \} ,\\
    \cB_{22} &=  \{ (x, y)\, | \, 0.6 \leq x \leq 0.72, 0.545 \leq y \leq 0.585 \} ,
\esplit
that enclose the second pair of islands; these are the blue boxes in \Fig{phase_portraits}(b).
Figure~\ref{fig:island2_trapping_hist.K1.L5} shows trapping time versus recurrence time histograms for $\cB_2 = \cB_{21} \cup \cB_{22}$. Once again, when noise is present, a population of trajectories spends increased time near the island chain, resulting in increased recurrence times. These histograms indicate that $\cB_2$ is responsible for some of the ``dust'' of points below the main tail seen in \Fig{trapping_hist.K1.L5}. However, the trapping within $\cB_2$ is less pronounced than that within $\cB_1$: there is less correlation between $t^R$ and $t^T$. We hypothesize that this is because, in order for a trajectory to move from $\Lambda$ to $\cB_2$, it must first cross the cantori that have replaced the rotational invariant circles of \Fig{phase_portraits}(a). By contrast, the region $\cB_1$ is on the ``same side" of the cantori as $\Lambda$. 
For the largest variance, shown in \Fig{island2_trapping_hist.K1.L5}(d), the distribution is more $t^T$-independent, even more pronounced than that seen in \Fig{trapping_hist.K1.L5}(d) for $\cB_1$.

\InsertFigFour{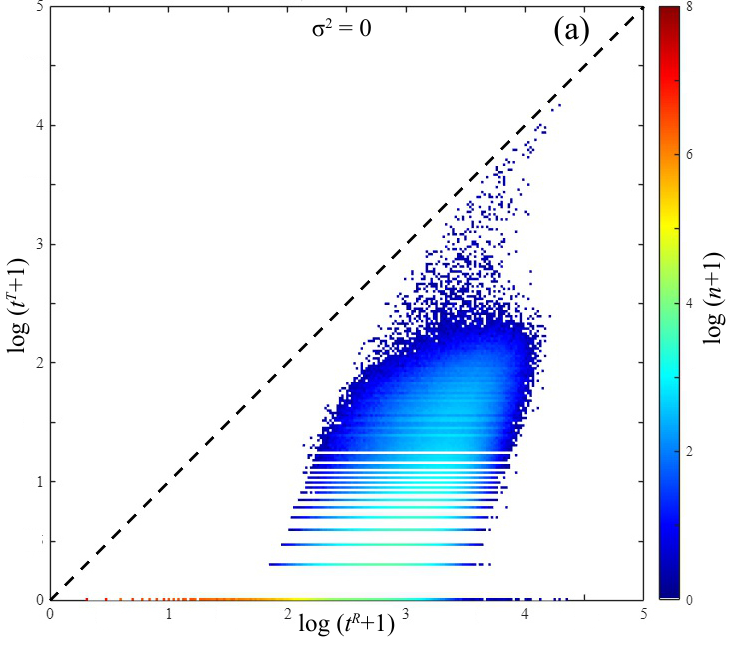}{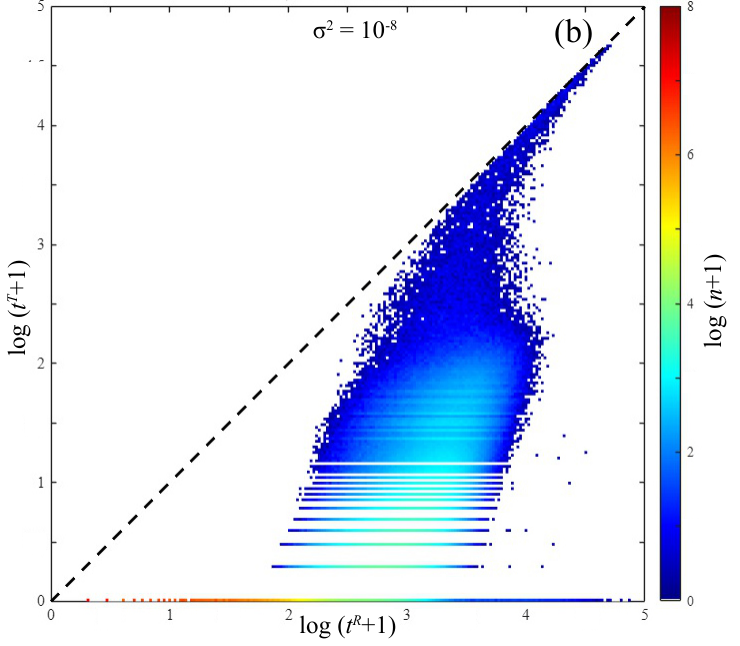}{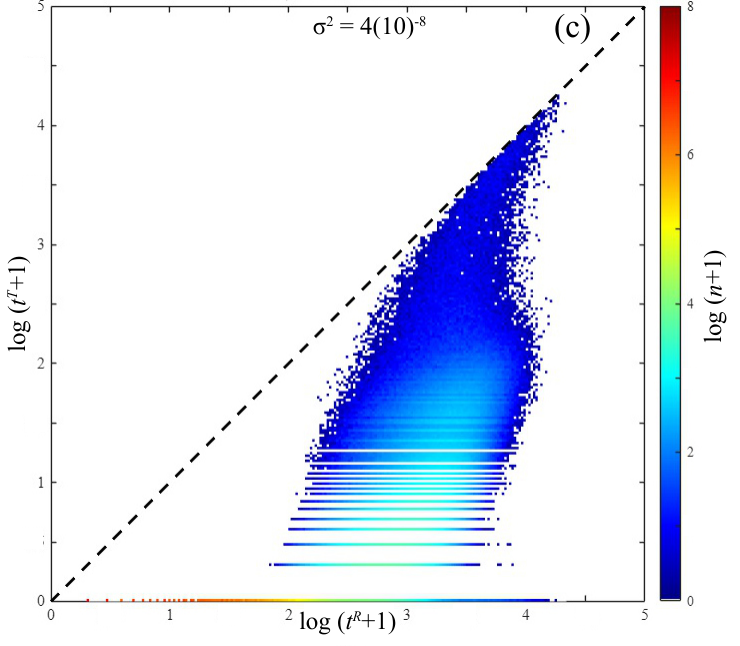}{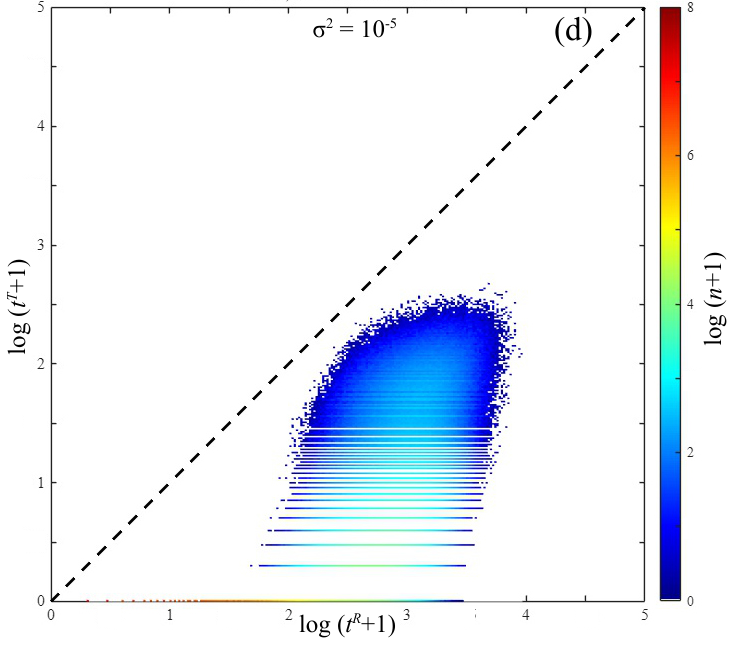}
{Histogram of trapping time in the region $\cB_2$ \Eq{SecondIsland} versus recurrence time to $\Lambda$ \Eq{StartBox} for the noisy Harper map with $(K,L)=(1,5)$. The dashed line indicates the maximum $t^T$ for a given $t^R$.}
{island2_trapping_hist.K1.L5}{3in}

\subsection{Phase Space Structures}

To visualize the phase space structure of orbits recurrent to $\Lambda$, \Fig{short_trapping} shows  trajectories that \textit{survive} up to $k = 100$, i.e., those with $t^R > 100$. Here we chose $10^5$ initial conditions distributed uniformly in $\Lambda$ for the nearly ergodic Harper map with $\sigma^2 =0$ in panel (a), and $\sigma^2 = 10^{-5}$ in panel (b). As we also saw in the PRS (\Fig{prs.K1.L5}), there are more trajectories ($4005$) that have not recurred at $t=k$ for $\sigma^2 = 10^{-5}$ than for the deterministic case ($2777$). 
Each point $(x_t,y_t)$ in the first $100$ iterates of a surviving orbit is colored by the current time $t$ as shown in the color bar. Note that the phase points for the largest times (yellow) of most of these surviving orbits are either near $\cB_1$ or within the region between the remnants of the rotational invariant circles. 
Thus, even though the cantori were not clearly visible in \Fig{phase_portraits}(b), we see that they have a strong influence on the surviving trajectories. 

\InsertFigTwo{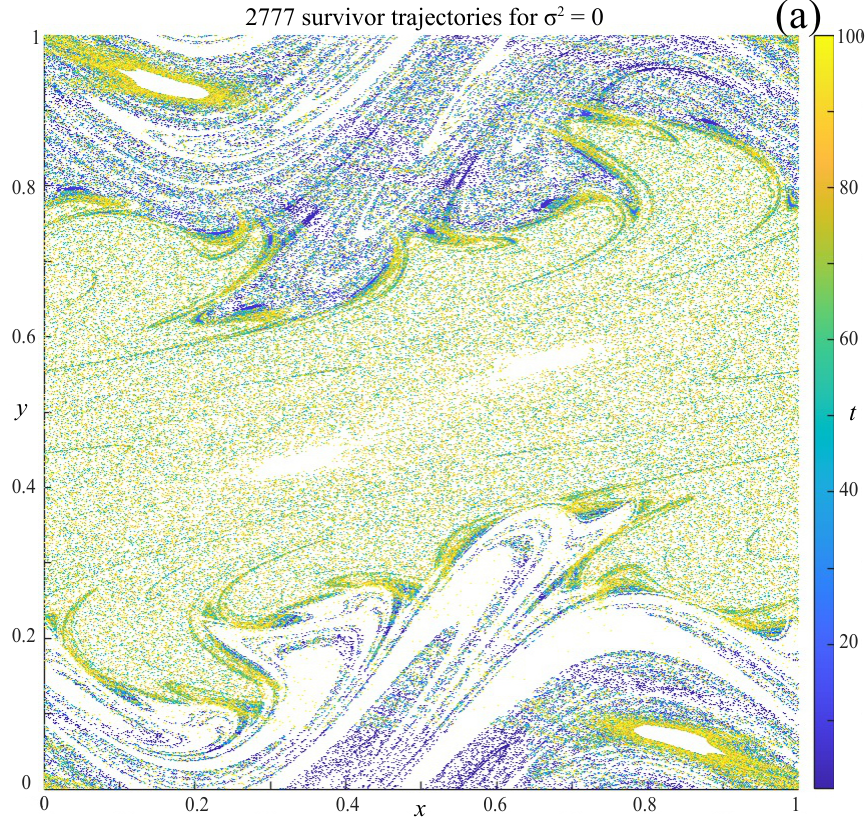}{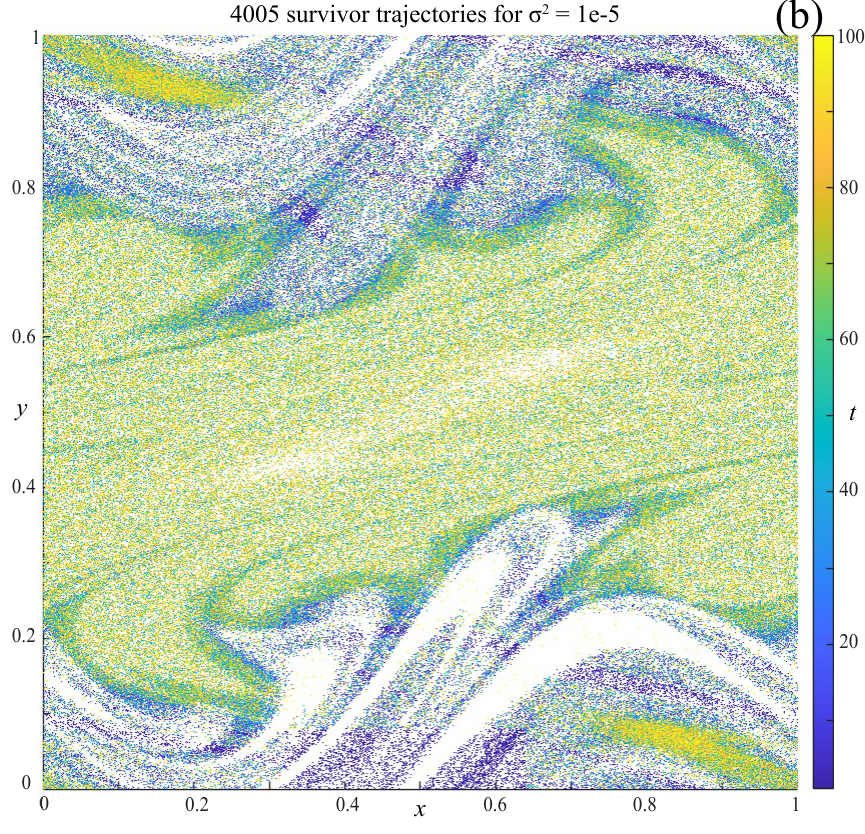}{Short-time survivors for $(K,L)=(1,5)$. The first $100$ iterates for trajectories with $t^R > k = 100$ for $10^5$ initial conditions uniformly distributed in $\Lambda$. (a) The $2777$ survivors for the deterministic case, and (b) the $4005$ survivors for Gaussian noise with $\sigma^2 = 10^{-5}$. Each iterate is colored by the current time along the orbit.}{short_trapping}{3in}

Figure~\ref{fig:long_trapping} shows phase portraits for orbits with the longest recurrence times, $t^R > k = 3(10)^4$. Here we plot the \textit{last} $100$ iterates of such trajectories selected from $10^8$ initial conditions in $\Lambda$. For the deterministic case in panel (a), only two trajectories survive: each is trapped at the edge of one of the period-two islands. For $\sigma^2 = 10^{-8}$ there are $102$ surviving trajectories as shown in panel (b); these spend most of their last $100$ iterates trapped \textit{inside} one of the island chains. 

\InsertFigTwo{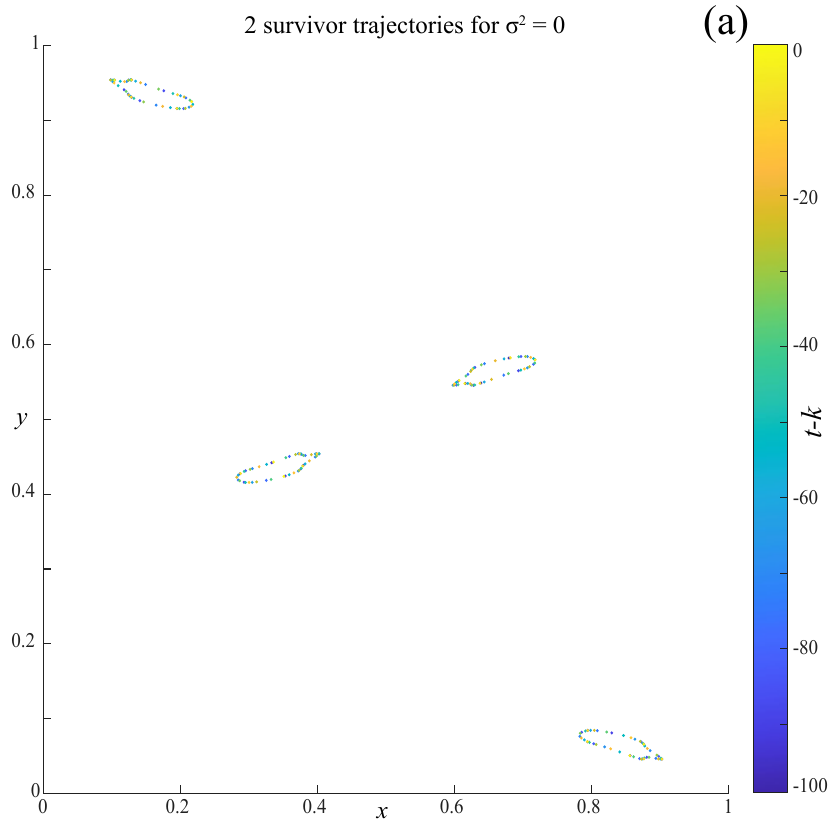}{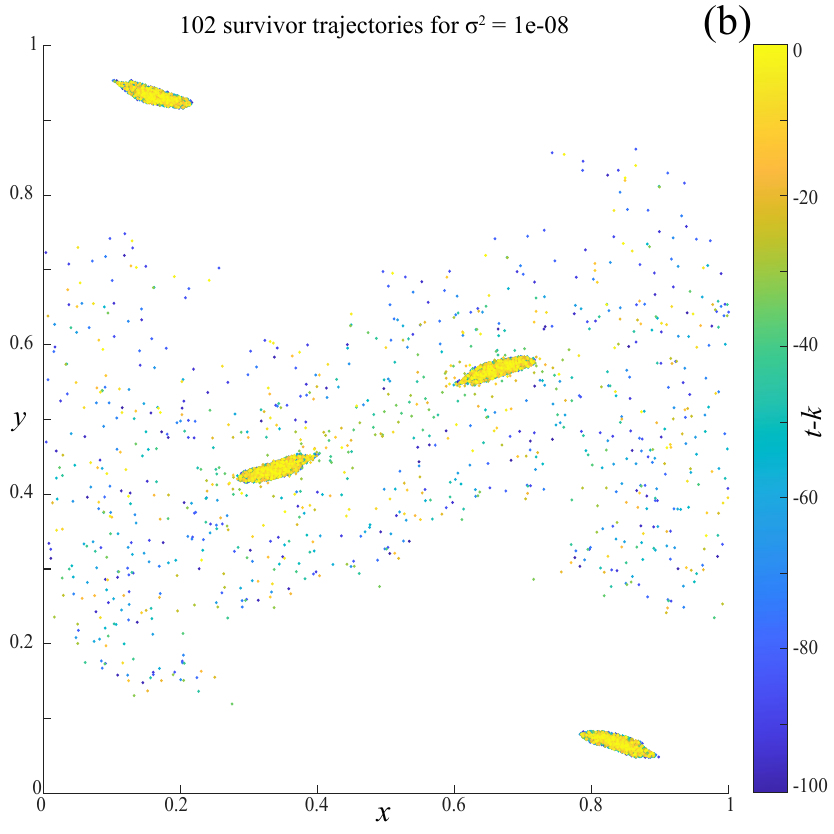}{Survivors of $N(0) = 10^8$ initial conditions in $\Lambda$ with $t^R > k = 3(10)^4$, for $(K,L)=(1,5)$. Shown are the final $100$ iterates before $t = k$. (a) The two survivors for the deterministic case, and (b) the $102$ survivors for Gaussian noise with $\sigma^2 = 10^{-8}$. Each point is colored by $t-k$, the number of steps before $k$.}
{long_trapping}{3in}

This shows how the stickiness of the islands correlates with the longest surviving trajectories for small, but nonzero, noise. Even though a noisy trajectory can just as easily diffuse out of an island as it can diffuse in, we see that when the noise level is small, the longest recurrence times are correlated with trapping.
The importance of islands is related to their size and the noise intensity.
For a purely random process with noise \Eq{Gaussian}, each component will undergo a random walk leading to a variance at time $t$ of 
\[
   \langle (\delta x(t))^2 \rangle = \langle (\delta y(t))^2 \rangle = \sigma^2 t, 
\]
since $\langle \delta x\rangle = \langle \delta y\rangle= 0$. 
Such a purely diffusive motion would lead to trajectories escaping islands of size $\rho$ in a time
of order
\beq{tescape}
    t_{escape} \sim \frac{\rho^2}{\sigma^2}
\eeq
In the nearly ergodic case, the island chains have scale size $\rho \approx 0.03$. For the variance $\sigma^2 = 10^{-8}$ of \Fig{long_trapping}(b), diffusive escape would occur at a time $t_{escape} \sim 9(10)^4 = 3k$, implying that the islands are still able to strongly influence trapping. It is worth noting that some of the surviving trajectories in \Fig{long_trapping}(b) are in the central chaotic region, outside of the island chains. These show the influence of the lower flux cantori in preventing recurrence to $\Lambda$. 

A similar analysis for $(K,L) = (0.6,4)$ is more complicated because there are many elliptic regions, as we saw in \Fig{phase_portraits}(a). Figure \ref{fig:survivors.K0.6L4} shows the surviving orbits with $t^R > k = 1000$, for four noise levels.
Note that as $\sigma$ increases the number of survivors grows
from $47$ for the deterministic case to $968$ for $\sigma = 10^{-2}$.
As we saw before, the regular regions exhibit stronger trapping for small, nonzero noise. For example, the period-two chain near $(x,y) = (0, 0.15)$ and $(0.5,0.15)$ traps more iterates for $\sigma^2 = 10^{-7}$ than does the deterministic case. However, when $\sigma^2 = 10^{-6}$ this chain is less distinct, 
and at $\sigma^2 = 10^{-4}$ it is no longer visible. For this noise level most of the surviving orbits have diffused across the bands of rotational invariant circles of the deterministic dynamics and are within the chaotic region $0.3 < y < 0.7$.

\InsertFigFour{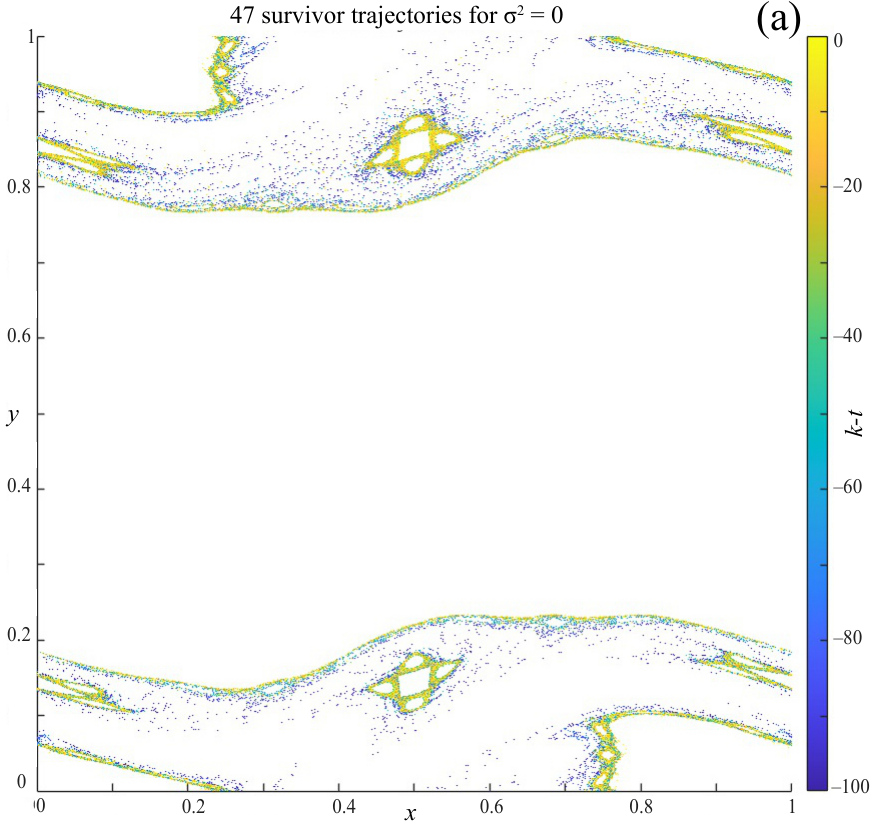}{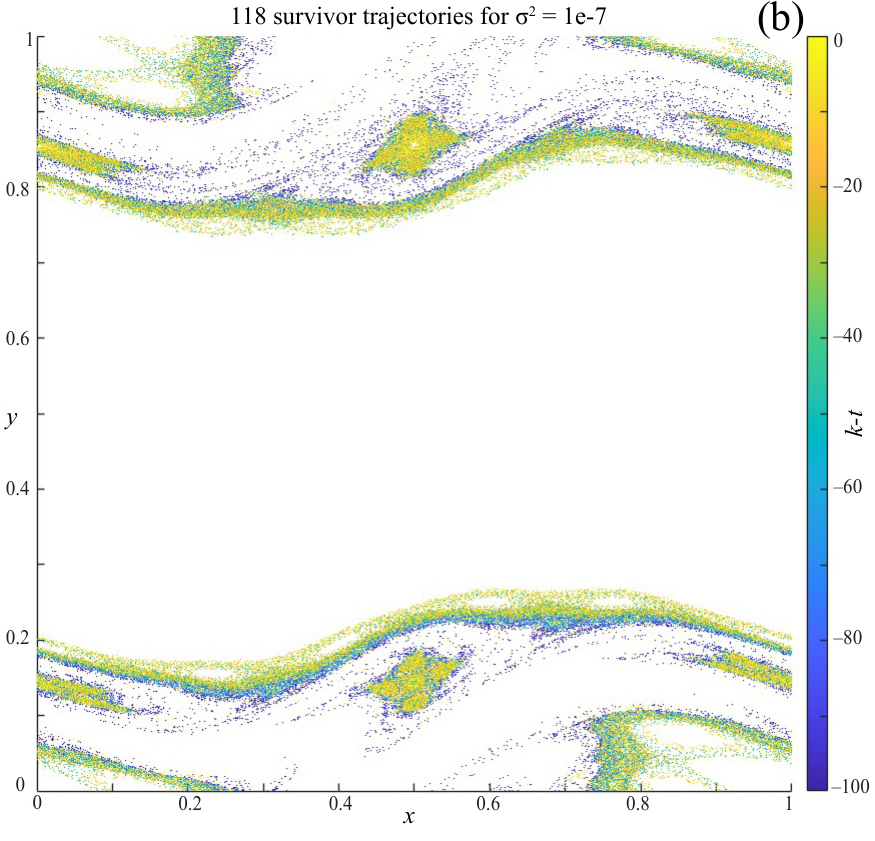}{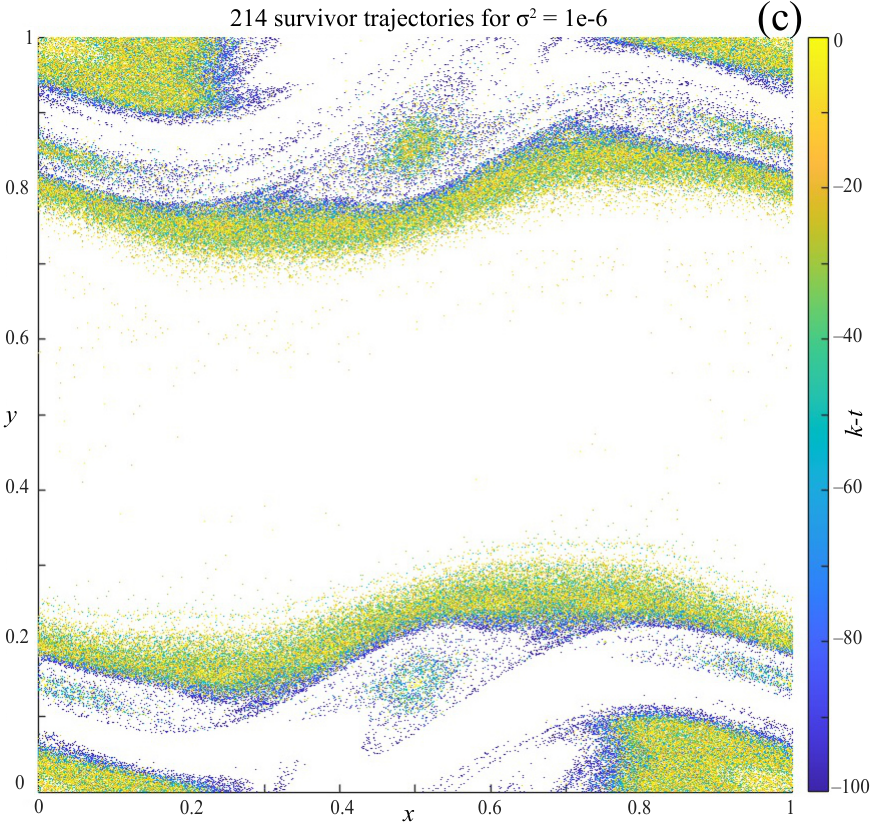}{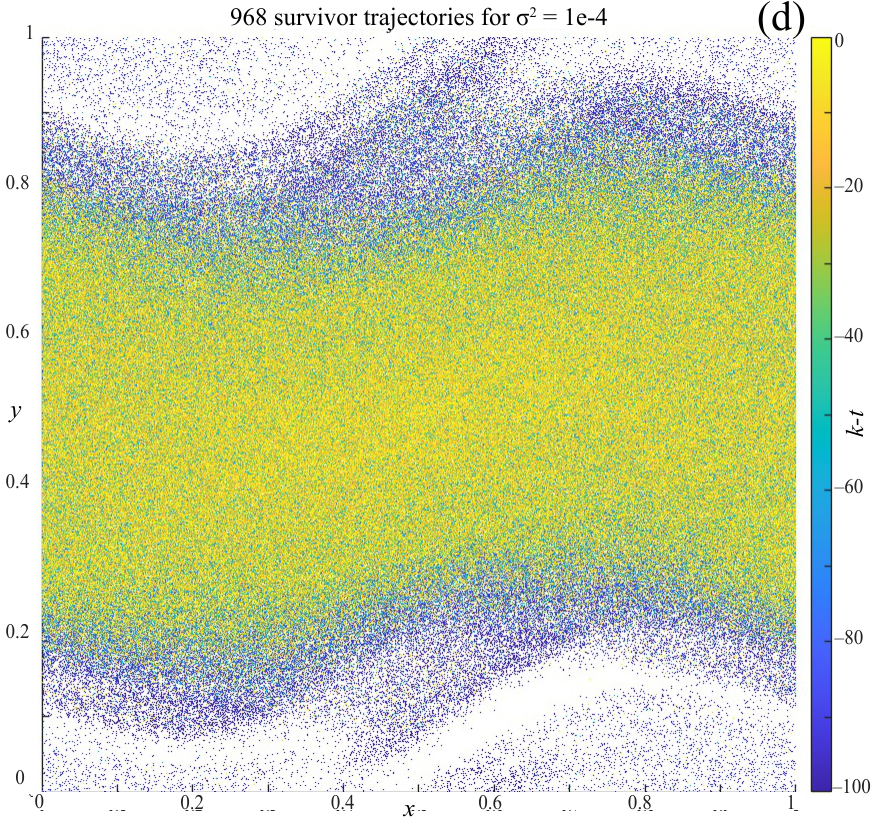}
{Survivors from $N(0) = 10^5$ initial conditions in $\Lambda$ with $t^R > k=1000$, for $(K,L) = (0.6,4)$. Noise intensities are indicated in each panel, and each point is colored by $t-k$.}
{survivors.K0.6L4}{3in}

 The period-two island chains in \Fig{survivors.K0.6L4} have size $\rho \approx 0.04$. Using the estimate \Eq{tescape}, we expect $t_{escape} \sim 2(10)^4$ for $\sigma^2 = 10^{-7}$, a long time on the scale of $k$. Consequently, trapping in these islands is prominent in \Fig{survivors.K0.6L4}(b). However, when $\sigma^2 = 10^{-4}$ in panel (d), $t_{escape} \sim 20$, implying---as seen in the figure---that these islands are not important at trapping orbits with $t^R > 1000$.

\section{Three-state Markov Model}\label{sec:ThreeState}
A simple explanation of the noise-induced stickiness observed in \Sec{Stickiness} can be obtained using a simplified, three-state Markov model of the dynamics. We suppose the phase space is divided into three regions $\cF, \cG$, and $\cB$ with areas $F, G$, and $B$, respectively. 
Furthermore, suppose that $F+G+B = 1$, and $F \approx G$ so that
\[
    0 \le B \ll F \le G.
\]
This is a simplified model of the nearly ergodic case of \Sec{NearlyErgodic}. We imagine that $\cF$ and $\cG$ correspond to large chaotic regions that are separated by relatively low-flux cantori, and $\cB$ represents an island chain, modeled here as a single invariant subset $\cB \subset \cF$. For this simple model, we consider the second island chain $\cB_2$ as part of the region $\cG$. Let $\Delta W$ denote the flux between $\cF$ and $\cG$ and $\Delta V$ be the noise-induced flux between $\cF$ and $\cB$, dependent on the intensity of the noise. We assume that
\[
    0 \leq \Delta V \leq B, \quad 0 \leq \Delta W \leq \min (F, G), 
\]
since the flux represents the portion of the region that escapes per step. A Markov model for this simplified system is
\beq{MarkovModel}
    \begin{pmatrix}
        \rho_F \\
        \rho_G \\
        \rho_B \\
    \end{pmatrix}'
     = 
     \begin{pmatrix}
         1 - \frac{\Delta W}{F} - \frac{\Delta V}{F} & \frac{\Delta W}{G} & \frac{\Delta V}{B} \\
         \frac{\Delta W}{F} & 1 - \frac{\Delta W}{G} & 0 \\
         \frac{\Delta V}{F} & 0 & 1 - \frac{\Delta V}{B} \\
     \end{pmatrix}
     \begin{pmatrix}
        \rho_F \\
        \rho_G \\
        \rho_B \\
     \end{pmatrix}.
\eeq
It is simple to confirm that 
\[
    \lambda_0 = 1, \quad
    \rho_0 = (F , G , B)^T ,
\]
is an eigenvalue-eigenvector pair that corresponds to a constant density equilibrium. The other two eigenvalues are
\beq{eigenvals}
    \lambda_{\pm} = \frac{1}{2} \left(2 - \alpha - \beta \pm \sqrt{\frac{F(\alpha - \beta)^2 + GB(\alpha + \beta)^2}{F + GB}} \right) ,
\eeq
where $\alpha = (\frac{1}{F} + \frac{1}{G}) \Delta W$, and $\beta = (\frac{1}{F} + \frac{1}{B}) \Delta V$. Note that under our assumptions, all eigenvalues are real. Let $\rho_\pm$ be the eigenvectors corresponding to the eigenvalues $\lambda_\pm$, respectively. 

For the deterministic case, $\Delta V = 0$: the island region $\cB$ is not accessible in the absence of noise. Then we have that $\lambda_+ = 1$, with $\rho_+ = (0,0,1)^T$, and $\cB$ is invariant. The third eigenpair is
\[
    \lambda_- = 1 - \alpha, \quad 
    \rho_- =   (1,-1,0)^T ,
\]
corresponding to mixing between $\cF$ and $\cG$. Thus for this model of the deterministic dynamics, the density decays to the uniform state at the rate $\lambda_-$.

In the presence of noise, $\Delta V > 0$ so that the island region $\cB$ is accessible. For this case the modulus of the second eigenvalue $\lambda_+$ will always be less than $1$, since $\alpha,\beta > 0$, and since $|\lambda_+| > |\lambda_-|$, $\rho_+$ decays more slowly than $\rho_-$. An example of the dependence of the eigenvalues on $\Delta V$ is shown in \Fig{MarkovEV}.

It is not hard to confirm that 
\[
    \frac{d \lambda_\pm}{d \beta} < 0
\]
for all $\beta$, so that both $\lambda_+$ and $\lambda_-$ are decreasing as $\Delta V$ increases. Since $\alpha < 2$ by our assumptions, we have that $1 = |\lambda_+| > |\lambda_-|$ when $\Delta V = 0$. 
By continuity, there must be some interval of $\beta$ such that $ 1> | \lambda_+| > |\lambda_-|$. In other words, the new slowest decaying eigenvector corresponds to $\lambda_+$ upon the introduction of noise. The implication is that the accessibility of an island due to noise results in a more slowly decaying tail for the PRS than for the deterministic case. This feature is also what we observed in \Fig{prs.K1.L5}: noise can result in long time tails.



\InsertFig{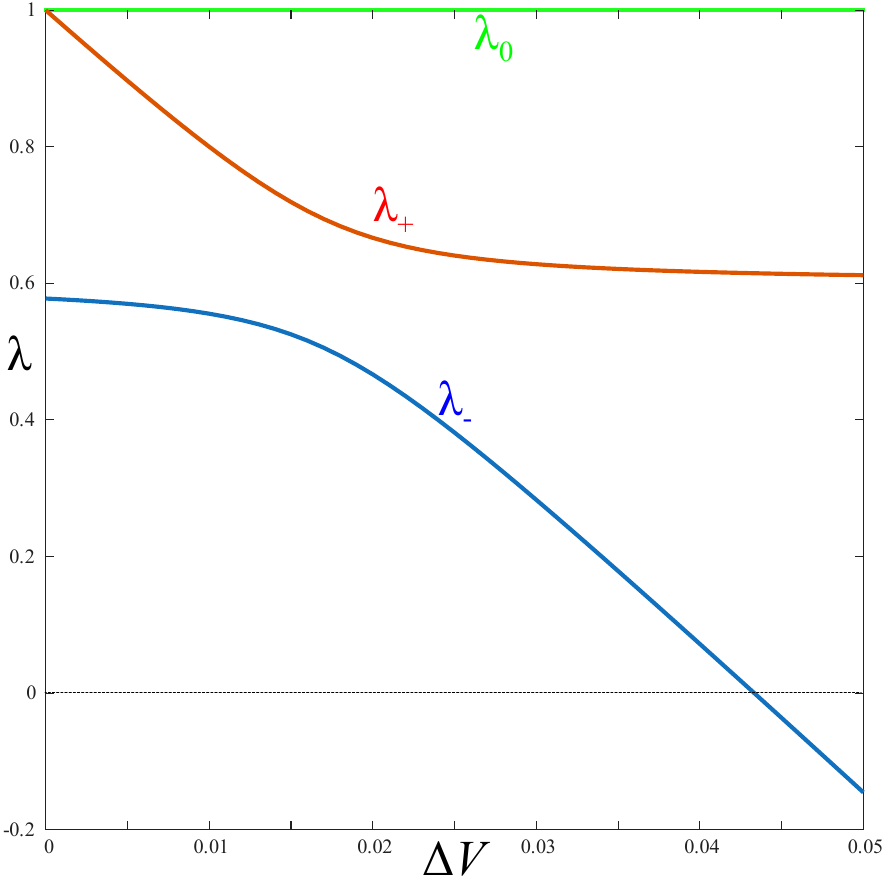}{Eigenvalues of the three-state Markov model as a function of the flux $\Delta V$ with $F = 0.45$, $G = 0.5$, $B = 0.05$, and $\Delta W = 0.1$.
}{MarkovEV}{2.5in}

\section{Conclusions}\label{sec:Conclusions}
We studied changes in the Poincar\'e recurrence statistic (PRS) for two representative parameter sets of the Harper map \Eq{HarperMap} with added noise from three distinct distributions. For $(K,L) = (0.6, 4)$, where the deterministic map has a mixture of chaotic and regular regions, the noisy PRS shows a short-time power-law decay with a slower rate than that a characteristic of the deterministic case. On long time scales, the PRS transitions to an exponential decay. For $(K,L) = (1,5)$ the deterministic dynamics has only four small island chains, and we find that these are responsible for long-term trapping in the noisy case: trajectories with long recurrence times tend to spend most of their iterates near the islands. When noise is added, trajectories can diffuse into the previous inaccessible islands, and this increases the recurrence times.

Even though a noisy trajectory can just as easily diffuse out of an island as it can diffuse in, when the noise level is small, the probability of leaving is also small. This leads to  tails in the recurrence time distribution as we saw in \Fig{prs.K1.L5}. However, when the noise intensity increases, random perturbations
cause a faster decrease in the PRS. Orbits spend long periods of time trapped near the island chains when the noise intensity is small, and the distribution becomes more uniform when the noise intensity increases.

When the phase space has a more generic mixture of regular and chaotic orbits, it is harder separate the effects of individual structures; however, plots of the survivor trajectories still demonstrate the effects of noise on trapping. A simple estimate based on a random walk confirmed that as long as the standard deviation from the deterministic trajectory is less than the size of the island, a noisy orbit will stay trapped by the island. 

The simple Markov model of this phenomenon in \Sec{ThreeState} showed that when an island becomes accessible due to noise, a new slowest decaying eigenvalue appears. This is consistent with the noise-induced slowing of the decay of the PRS. This new eigenvalue decreases from one as the noise intensity grows. 
The results of the Markov model could be extended to more states:  this is the idea behind the Ulam method, \cite{Froyland99} which approximates the Peron-Frobenius operator by discretizing the phase space. However, such a finite discretization cannot fully respect the fractal nature of the invariant subsets of the true dynamics, nor will it be able to show the power-law decay expected for the deterministic case.  

In the future we plan to study the effect of noise on higher-dimensional volume-preserving dynamics, such as the ABC map. Such systems can model mixing of passive scalars in an incompressible flow.

\begin{acknowledgments}
Support from a gift from the Northrop-Grumman University Research fund is gratefully acknowledged.
\end{acknowledgments}

\bibliography{NoiseRefs}

\end{document}